# STELLAR ORIGINS OF EXTREMELY $^{13}$C- AND $^{15}$N-ENRICHED PRESOLAR SIC GRAINS: NOVAE OR SUPERNOVAE?

Nan Liu[1], Larry R. Nittler[1], Conel M. O'D. Alexander[1], Jianhua Wang[1]

Marco Pignatari[2, 7, 8], Jordi José[3,4] & Ann Nguyen[5,6]

[1]Department of Terrestrial Magnetism, Carnegie Institution for Science,

Washington, DC 20015, USA nliu@carnegiescience.edu;

[2] **E.A. Milne Centre for Astrophysics, Department of Physics & Mathematics, University of Hull, HU6 7RX, UK;**

[3]**Department de Fisica, EUETIB, Universitat Politècnica de Catalunya,**

**E-08036 Barcelona, Spain;**

[4]Institut d'Estudis Espacials de Catalunya, E-08034 Barcelona, Spain;

[5]Robert M. Walker Laboratory for Space Science, Astromaterials Research and Exploration Science Directorate, NASA Johnson Space Center, Houston, TX 77058, USA;

[6]Jacobs, NASA Johnson Space Center, Houston, TX 77058, USA;

[7] **Konkoly Observatory, Research Centre for Astronomy and Earth Sciences, Hungarian Academy of Sciences, Konkoly Thege Miklos ut 15-17, H-1121 Budapest, Hungary.**

ABSTRACT

Extreme excesses of $^{13}$C ($^{12}$C/$^{13}$C<10) and $^{15}$N ($^{14}$N/$^{15}$N<20) in rare presolar SiC grains have been considered diagnostic of an origin in classical novae, though an origin in core collapse supernovae (CCSNe) has also been proposed. We report C, N, and Si isotope data for 14 submicron- to micron-sized $^{13}$C- and $^{15}$N-enriched presolar SiC grains ($^{12}$C/$^{13}$C<16 and $^{14}$N/$^{15}$N<~100) from Murchison, and their correlated Mg-Al, S, and Ca-Ti isotope data when available. These grains are enriched in $^{13}$C and $^{15}$N, but with quite diverse Si isotopic signatures. Four grains with $^{29,30}$Si excesses similar to those of type C SiC grains likely came from CCSNe, which experienced explosive H burning occurred during explosions. The independent coexistence of proton- and neutron-capture isotopic signatures in these grains strongly supports heterogeneous H ingestion into the He shell in pre-supernovae. Two of the seven putative nova grains with $^{30}$Si excesses and $^{29}$Si depletions show lower-than-solar $^{34}$S/$^{32}$S ratios that cannot be explained by classical nova nucleosynthetic models. We discuss these signatures within the CCSN scenario. For the remaining five putative nova grains, both nova and supernova origins are viable because explosive H burning in the two stellar sites could result in quite similar proton-capture isotopic signatures. Three of the grains are sub-type AB grains that are also $^{13}$C enriched, but have a range of higher $^{14}$N/$^{15}$N. We found that $^{15}$N-enriched AB grains (~50<$^{14}$N/$^{15}$N<~100) have distinctive isotopic signatures compared to putative nova grains, such as higher $^{14}$N/$^{15}$N, lower $^{26}$Al/$^{27}$Al, and lack of $^{30}$Si excess, indicating weaker proton-capture nucleosynthetic environments.

---

[8]NuGrid collaboration, http://www.nugridstars.org.

1. INTRODUCTION

Primitive meteorites contain several types of dust grains that condensed in stellar winds or in the ejecta accompanying stellar explosions, became part of the protosolar molecular cloud, and survived destruction in the early Solar System before incorporation into meteorites. These presolar grains are identified by their anomalous isotopic compositions, which cannot be explained by any known mass fractionation process in the Solar System and instead suggest different stellar origins (Zinner et al. 2014). The mineral silicon carbide (SiC) is the best studied presolar grain phase because (1) it can be quite easily separated from bulk meteorites by acid dissolution (Amari et al. 1994); and (2) SiC is predominantly formed in C-rich gas according to grain condensation calculations in stellar conditions (Lodders et al. 1995; Ebel & Grossman 2001), so its formation in the generally O-rich (O/C>1) Solar System is rare; (3) Compared to other presolar mineral phases such as oxides (typically submicron in size), many SiC grains are relatively larger so that isotopic analysis of multiple elements can be made in each single grain.

Based on the measured isotopic compositions of a large number of elements, more than 90% of presolar SiC (mainstream grains) have been inferred to originate in winds from low-mass asymptotic giant branch (AGB) stars with close-to-solar metallicity (Alexander 1997; Nittler 2003; Davis 2009; Zinner 2014). Mainstream grains have $^{12}$C/$^{13}$C between 10 and 100 with a wide range of $^{14}$N/$^{15}$N ratios; their δ$^{29}$Si/$^{28}$Si[9] and δ$^{30}$Si/$^{28}$Si vary from −100‰ to 200‰ (see Zinner 2014 for details). The remaining 10% of presolar grains have been divided into isotopic sub-groups, including X and C grains from core collapse supernovae (CCSNe), Y and Z grains from low metallicity AGB stars, and AB grains, which possibly have multiple stellar sources (e.g., Alexander 1993; Nittler et al. 1996; Hoppe et al. 1997; Amari et al. 1999, 2001a; Pignatari et al. 2015, hereafter P15). Relative to mainstream grains, Y and Z grains are more enriched in $^{30}$Si and additionally, Y grains have $^{12}$C/$^{13}$C ratios higher than 100. Although X and C grains both came from CCSNe with similar C and N isotopic compositions, X grains are enriched in $^{28}$Si (negative δ$^{29,30}$Si/$^{28}$Si), while C grains show excesses in both $^{29}$Si and $^{30}$Si (positive δ$^{29,30}$Si/$^{28}$Si). Finally, AB grains are mainly characterized by $^{12}$C/$^{13}$C ratios lower than 10, and their Si isotopic compositions are overlapped with those of mainstream grains (see Zinner 2014 for details).

A small fraction of presolar SiC grains are characterized by extremely low $^{12}$C/$^{13}$C and $^{14}$N/$^{15}$N ratios and excesses in $^{30}$Si relative to $^{29}$Si (Amari et al. 2001b). Such isotopic signatures agree well with the characteristic proton-capture nucleosynthetic signatures in novae. Novae are powered by thermonuclear explosions in the H-rich envelopes of white dwarf (WD) stars. These H-rich envelopes are produced by transfer of material from a low-mass stellar companion that is still on the main sequence or on the giant branch. There are two types of classical novae, depending on the nature of the WD: CO WDs are the remnants of evolved AGB stars less massive than ~6−8 $M_\odot$, and ONe WDs are remnants of more massive AGB stars (8−10 $M_\odot$, José et al. 2004; Herwig 2005; Jones et al. 2013; Karakas & Lattanzio 2014). Nucleosynthetic models predict ubiquitous production of $^{13}$C and $^{15}$N by explosive H burning in both CO ($M > 0.8\ M_\odot$) and ONe novae (José & Hernanz 2007).

These rare presolar SiC grains are therefore called putative nova grains. A number of problems, however, are faced by nova models in explaining the grain data: (1) putative nova grains of different mineral phases including SiC, oxides, and silicates, all have much less anomalous isotopic compositions compared to nova ejecta that therefore needs to be highly diluted with isotopically normal (i.e., ~solar) material (>90%) to match the grain data (Amari et al. 2001b; Nittler & Hoppe 2005; Gyngard et al. 2010; Nittler et al. 1997; Leitner et al. 2012; Nguyen & Messenger 2014); (2) although thermodynamic equilibrium calculations predict that SiC grains can condense in the innermost ejected shell of ONe novae (José et al. 2004), mixing with a large amount of solar-like composition material (C<O) would greatly lower the C/O ratio in nova ejecta, and consequently lower the possibility for SiC to condense during nova outbursts; (3) CO novae are more abundant (70%–80%), and also more efficient in producing dust than ONe novae (e.g., Gehrz et al. 1986, Mason et al. 1996), but most of the putative nova grains seem to be from ONe novae based on their Si isotope data (Amari et al. 2001b). These difficulties therefore make it quite problematic to link $^{13}$C- and $^{15}$N-enriched grains to nova origins. On the other hand, it is important to point out that both C- and O-rich dust has been simultaneously observed in some novae (e.g., Gehrz et al. 1998). Thus, the strong radiation from the underlying WD may cause the process of condensation to proceed kinetically rather than at equilibrium. Consequently, the strong role of the CO molecule in determining formation of C-rich dust in reduced stellar environments (C/O>1) could be dramatically decreased (Shore & Gehrz 2004).

---

[9]δ-values are deviations from normal isotopic ratios in parts per thousand,
$$\delta^{29}\text{Si}/^{28}\text{Si} = [\frac{(^{29}\text{Si}/^{28}\text{Si})_{grain}}{(^{29}\text{Si}/^{28}\text{Si})_{standard}} - 1] \times 1000 \ .$$

Interestingly, Nittler & Hoppe (2005) reported one $^{13}$C- and $^{15}$N-enriched SiC grain (M11-334-2) with excesses in $^{28}$Si, $^{44}$Ca, and $^{49}$Ti that point towards a CCSN origin. Consequently, this result suggests that the proton-capture nucleosynthesis associated with classical novae may also occur in some CCSNe, perhaps due to ingestion of H from the stellar envelope into underlying layers (*e.g.*, P15). To better resolve the stellar origin(s) of putative nova grains, isotope data of a number of elements in single grains are needed to constrain the nucleosynthetic processes in their parent stars. Nova models predict that $^{26}$Al/$^{27}$Al ratios of nova ejecta are generally lower than the initial $^{26}$Al/$^{27}$Al ratios typically measured in X grains from CCSNe and that Ti lies beyond the standard nucleosynthetic endpoint for classical nova outbursts, i.e., Ti isotopic compositions of nova ejecta should be close to those of mainstream grains in most cases (José et al. 2004). In contrast, $^{49}$Ti and $^{50}$Ti excesses, relative to $^{48}$Ti, were found in a majority of X grains (Nittler et al. 1996; Lin et al. 2010). Thus, the initial $^{26}$Al/$^{27}$Al and Ti isotope ratios can be used as diagnostic tools to distinguish CCSNe from novae. Limited by their extremely low abundances (<1%) and small sizes (<1 μm), most of the fewer than 10 putative nova grains in the literature were only analyzed for the isotopic compositions of their C, N, and Si (Amari et al. 2001b). Magnesium-Al isotope ratios were only reported for four grains, and Ti isotope ratios for two grains, both of which had Ti isotope anomalies contradicting with a nova origin (Amari et al. 2001b; Nittler & Hoppe 2005).

In this paper, we report C, N, and Si isotope data for 11 new putative nova grains and three $^{15}$N-enriched AB grains from Murchison and their correlated Mg-Al, S, and Ca-Ti isotope data when available. These new $^{13}$C- and $^{15}$N-enriched grains double the number of putative nova grains found in meteorites. More importantly, they provide valuable information on nucleosynthesis in their parent stars through isotope data of heavier elements (*e.g.*, S, Ti). While seven of the 11 new putative nova grains are enriched in $^{30}$Si relative to $^{29}$Si, as seen in previous nova grains (Amari et al., 2001b), four grains show enrichments in both $^{29}$Si and $^{30}$Si, relative to $^{28}$Si, by up to 15 times their solar abundances, which are quite similar to the Si isotopic compositions of C grains that are believed to have originated in CCSNe (Hoppe et al. 2005; Xu et al. 2015; Pignatari et al. 2013a). By definition, these grains cannot be classified as either putative nova grains or C grains. Historically, presolar SiC grains with unusual C, N, and Si isotope ratios that could not be assigned to any existing groups have often been named type U (unusual or ungrouped) grains. The definition of U grains, however, is quite ambiguous, and not all grains classified as U-type in the recent literature may be genetically related. As will be discussed in Section 4.1, $^{13}$C- and $^{15}$N-enriched grains with excesses in $^{29,30}$Si essentially originated from similar regions in CCSNe as type C grains. We therefore propose to rename type C grains ($^{12}$C/$^{13}$C > 10) "*C1 grains*" and name the $^{13}$C- and $^{15}$N-enriched grains ($^{12}$C/$^{13}$C < 10) with $^{29,30}$Si excesses "*C2 grains*". By comparing with recent nucleosynthetic calculations by P15, the C2 grains from this study confirm the possibility raised by Nittler & Hoppe (2005) that some, if not all, of $^{13}$C- and $^{15}$N-enriched presolar SiC grains are from CCSNe.

2. Experimental Methods

The SiC grains in this study were extracted from the Murchison meteorite by means of the CsF isolation method described by Nittler & Alexander (2003). The grains were separated in size by sedimentation. Thousands of grains with typical diameter ~1 μm were dispersed on a high purity Au foil from a water suspension and further pressed into the Au with a flat sapphire disk.

We searched for rare SiC grains by manually acquiring simultaneous ion images of $^{12}$C$_2^-$, $^{12}$C$^{13}$C$^-$, $^{12}$C$^{14}$N$^-$, $^{12}$C$^{15}$N$^-$, $^{28}$Si$^-$, $^{29}$Si$^-$, and $^{30}$Si$^-$ in multi-collection mode using a focused Cs$^+$ ion beam (1−2 pA, 100−150 nm) with a Cameca NanoSIMS 50L ion microprobe at the Carnegie

Institution for Science. The use of $C_2$ rather than C ions allows (1) better alignment of the C and N isotopic secondary beams in the mass spectrometer because of similar energy distributions (differing energy distributions between **atomic** and molecular ions, Gregorio et al. 2013); and (2) better alignment of the C and Si isotopic secondary beams because of smaller differences in mass. Data were acquired in imaging mode and isotope ratios for individual grains determined with the L'image processing software written by Dr. L. R. Nittler. Aggregates of synthetic SiC and $Si_3N_4$ grains were used as isotopic standards and measured in between every 10−20 grain measurements (~5 to 10 min integration time for each grain analysis) in order to monitor and correct for instrumental drift in the isotopic mass fractionations. After NanoSIMS analysis, we obtained high-resolution images (~5 nm/pixel) and energy-dispersive x-ray spectra of candidate SiC grains with a JEOL 6500F field-emission SEM. The SEM images showed that some SiC grains that appear to be single grains in NanoSIMS ion images are in fact clumps of smaller SiC grains (< 1 μm). Thus, there exist large uncertainties in estimating the total number of single SiC grains analyzed only based on the NanoSIMS ion images, making it difficult to accurately ascertain SiC sub-type abundances in the Murchison acid residue.

Out of ~700 ion images collected on more than 1000 grains, we identified 10 SiC grains that are extremely enriched in $^{13}C$ and $^{15}N$ ($^{12}C/^{13}C<16$ and $^{14}N/^{15}N<50$) and three AB grains. One of the 13 grains was only 200 nm across and sputtered away during the analysis of C, N, and Si isotopes. We measured S isotopes in four of the 12 remaining grains. These measurements were made by simultaneously obtaining ion images of $^{12}C_2^-$, $^{12}C^{13}C^-$, $^{12}C^{14}N^-$, $^{12}C^{15}N^-$, $^{32}S^-$, $^{33}S^-$, and $^{34}S^-$ in three of the four grains, and ion images of $^{12}C_2^-$, $^{28}Si^-$, $^{29}Si^-$, $^{30}Si^-$, $^{32}S^-$, $^{33}S^-$, and $^{34}S^-$ in grain GAB in multi-collection mode by rastering a $Cs^+$ beam over the grain areas. Grains G270_2 and Ag2_6 were completely sputtered away during S isotope analysis. Sulfur contamination in the synthetic SiC standard is high, so it was used as a standard for the S analysis. Also, because the S isotope ratios of mainstream grains are quite normal according to previous studies (*e.g.*, Hoppe et al. 2015), we also measured mainstream grains adjacent to the four grains to monitor the instrumental isotope mass fractionation. The S concentrations of the mainstream grains ($^{32}S/^{28}Si$ ion ratios in the range of 0.007 to 0.07) are 10−100 times lower than that of the SiC standard ($^{32}S/^{28}Si$ ion ratios of ~0.7). Overall, the $\delta^{33}S/^{32}S$ and $\delta^{34}S/^{32}S$ values of all the mainstream grains are solar within uncertainties (~20‰), when normalized to the mean ratios of the synthetic SiC standard.

Magnesium isotopes were measured with Al by simultaneously collecting $^{24}Mg^+$, $^{25}Mg^+$, $^{26}Mg^+$, $^{27}Al^+$, $^{28}Si^+$, $^{30}Si^+$, and $^{48}Ti^+$ in the 10 remaining grains in multi-collection mode by rastering a primary O¯ beam (4 pA, ~300 nm). Subsequently, the 10 grains were also measured for $^{28}Si^+$, $^{40}Ca^+$, $^{44}Ca^+$, $^{47}Ti^+$, $^{48}Ti^+$, $^{49}Ti^+$, $^{50}Ti^+$, and $^{50}Cr^+$ in combined-analysis mode (Set 1: $^{28}Si^+$; Set 2: the rest of the isotopes). Seven of the 10 grains had detectable Ti counts that were correlated with Si in the ion images. Burma spinel was used as a standard for Mg isotope analysis and also used to determine the Mg and Al relative sensitivity factors, giving a relationship between secondary ion yields and the atomic ratios of $Al^+/Mg^+ = 1.16 \times Al/Mg$. The TiC standard was found to contain significant Ca contamination and was thus used as standard for both Ca and Ti isotope analysis. All the isotope data are reported in Table 1 with 1σ uncertainties. Note that an additional $^{13}C$- and $^{15}N$-enriched SiC grain from Murchison, G240-1, previously reported in an abstract by Nittler et al. (2006), is also discussed here and included in Table 1. This grain was found by automated C and Si isotopic measurements of 1550 Murchison SiC grains with the Carnegie ims-6f ion probe (Nittler & Alexander 2003) and re-measured for C, N, Si, and Al-Mg by NanoSIMS, using similar methods to those described above.

## 3. RESULTS

### *3.1 Putative Nova and $^{15}N$-enriched AB Grains*

Presolar SiC grains that are highly enriched in $^{13}C$ ($^{12}C/^{13}C<10$) are mainly assigned to one of two sub-types: AB and putative nova grains. AB grains have a wide range of $^{14}N/^{15}N$ ratios from ~50 to ~$10^4$ (solar $^{14}N/^{15}N$ ratio: 441±5, Marty et al. 2011), and Si isotopic ratios that are similar to those of mainstream SiC grains. The Si isotopes of mainstream SiC grains ($\delta^{29}Si/^{28}Si =(1.37\pm0.01)\times\delta^{30}Si/^{28}Si+(-19.9\pm0.6)$, Zinner et al. 2007) are roughly consistent with the predictions of Galactic chemical evolution (GCE, Lugaro et al. 1999; Zinner et al. 2006). On the other hand, putative nova grains are more enriched in $^{15}N$ and have distinctive Si and Al-Mg isotopes (*e.g.*, Fig. 2 of Zinner 2014). In Fig. 1, three new $^{15}N$-enriched AB grains and seven new putative nova grains from this study are plotted along with literature data for C, N, Si, and Al isotope ratios. Two grains (M11-334-2 and M11-151-4) reported by Nittler & Hoppe (2005) with Ti and/or Si isotopic compositions that are incompatible (*e.g.*, $^{28}Si$, $^{44}Ca$ excesses in M11-334-2; $^{47}Ti$ excess in M11-151-4) with nova nucleosynthesis (e.g., $^{30}Si$ excess and Ti isotope ratios close to mainstream grain data) are labeled in the plot. In addition, these two grains have the highest $^{26}$Al/$^{27}$Al ratios among all $^{13}$C- and $^{15}$N-enriched presolar SiC grains. Because of these distinctive isotopic signatures, we do not consider them as putative nova grains hereafter.

Figure 1 shows that the isotopic compositions of the seven new putative nova grains and three AB grains from this study are similar to previous ones for C, N, and Al. Interestingly, putative nova grains show higher initial $^{26}$Al/$^{27}$Al ratios than AB grains, but lower than X grains (X grains: 0.1<$^{26}$Al/$^{27}$Al<1, *e.g.*, Zinner 2014). Grain 577 is classified as an AB grain in the presolar grain database (Hynes & Gyngard 2007), whilst Huss et al. (1997) argued that G577 was from a low metallicity AGB star that had experienced very extensive deep mixing. In fact, although the $^{14}$N/$^{15}$N ratio of G577 is in the range of AB grains, its $^{26}$Al/$^{27}$Al and Si isotope ratios are more like those of the putative nova grains and it should probably be thus classified as such. Nitrogen-15 enriched AB grains and putative nova grains bear some similarity in that most grains of each type have solar Ti isotopic compositions and no excess in $^{44}$Ca (ascribed to decay of $^{44}$Ti, with a half-life of 60 yr, in grains from CCSNe), and that a few grains of each type show negative $\delta^{34}$S/$^{32}$S values (Fujiya et al. 2013). Putative nova grains, however, can be distinguished from $^{15}$N-enriched AB grains by (1) relatively lower $^{14}$N/$^{15}$N ratios (< 70); (2) higher $^{26}$Al/$^{27}$Al ratios (> 0.01); and (3) excesses in $^{30}$Si relative to $^{29}$Si.

*3.2 Al Contamination in Presolar SiC Grains*

In previous studies, surface contamination has been found in presolar SiC grains for both major (*e.g.*, C, N, Nittler & Alexander 2003) and trace elements (*e.g.*, Sr, Ba, Liu et al. 2015). The most common practices to remove surface contamination include: (1) acid-cleaning grains prior to isotopic and elemental analysis (Liu et al. 2014, 2015); and (2) surface-cleaning of grains by removing a few atomic layers using the Cs$^+$ ion source during NanoSIMS analysis. Groopman et al. (2015), however, found that $^{26}$Mg/$^{24}$Mg and $^{27}$Al/$^{24}$Mg ratios are well correlated in depth profiles of many presolar SiC grains, and the obtained linear fits often have negative $^{26}$Mg/$^{24}$Mg intercepts that are unphysical, indicating that there exists constant Al contamination (up to 60%) throughout NanoSIMS measurements (see Groopman et al. 2015 for calculations in detail). Thus, the "true" initial $^{26}$Al/$^{27}$Al ratios should be 1.5−2 times higher on average than the values calculated using the standard approach (e.g., equation (2) of Nittler et al. 1997). We also examined the depth profiles of Mg-Al isotope data for the ten grains reported in Table 1 and found similar "internal isochrones". These $^{13}$C- and $^{15}$N-enriched SiC grains, however, have extremely high $^{27}$Al/$^{24}$Mg ratios (>500 in most of the cases), so the intercepts of the linear fits have large uncertainties, and the amounts of Al contamination therefore cannot be accurately determined for these grains using the method of Groopman et al. (2015).

The grain GAB has the lowest $^{27}$Al/$^{24}$Mg of the measured grains and its depth profile is shown in Fig. 2 for illustration. During the first five cycles, because of surface Al contamination and variation in the Al$^+$/ Mg$^+$ ion ratio, the corresponding data points (grey circles) deviate from the linear fit (solid black, with 95% confidence bands shown as grey areas) to the data points that reached a steady state Al$^+$/$^{24}$Mg$^+$ ion ratio (red circles). Despite large uncertainties, Fig. 2b shows that the intercept is slightly negative, indicating that there might be $20^{+60}_{-15}$ % constant Al contamination, considering the 95% uncertainties in the linear fit. Because of the large uncertainties in the estimated amounts of Al contamination, the effect of Al contamination is not included in the initial $^{26}$Al/$^{27}$Al ratios reported in Table 1 and these thus represent lower limits. The relative difference in $^{26}$Al/$^{27}$Al ratios between putative nova and AB grains, however, cannot be explained by Al contamination, because $^{26}$Al/$^{27}$Al ratios of all putative nova grains are higher than those of $^{15}$N-enriched AB grains despite different instruments, samples, and analysis conditions used in these studies.

## 4. DISCUSSION

### 4.1 Stellar Origin of Type C2 Grains: CCSNe with Explosive H Burning

#### 4.1.1 Nova Models Compared to Type C2 Grains

Carbon-13 is produced in stars by proton capture on preexisting $^{12}$C followed by β-decay, $^{12}$C$(p,\gamma)^{13}$N$(\beta^+v)^{13}$C, which occurs during H burning in the CNO cycle at temperatures ranging from 1.5×10$^7$ to 3.5×10$^8$ K. Although the $^{13}$C production is enhanced with increasing temperature, $^{13}$C is also more efficiently consumed via $^{13}$N$(p,\gamma)^{14}$O$(\beta^+v)^{14}$N at sufficiently high T during the nova thermal nuclear runaway (TNR). As a consequence, the $^{12}$C/$^{13}$C production ratio does not depend strongly on temperature. On the other hand, $^{15}$N can only be significantly produced at high temperatures ($^{14}$N/$^{15}$N < 0.1). Thus, hot H burning during the CNO cycle (>10$^8$ K) is needed in order to produce significant amounts of $^{13}$C and $^{15}$N (José et al. 2004). Amari et al. (2001b) discussed difficulties in explaining both low $^{12}$C/$^{13}$C and $^{14}$N/$^{15}$N ratios of putative nova grains by nucleosynthesis in AGB stars, CCSNe, and Wolf-Rayet stars, which led to the conclusion that novae, ONe novae in particular, are the most likely stellar source of highly $^{13}$C- and $^{15}$N-enriched presolar SiC grains.

The only $^{13}$C- and $^{15}$N-enriched stellar site identified by both astrophysical observations (*e.g.*, Sneden & Lambert 1975) and nucleosynthetic simulations (*e.g.*, José et al. 2004, Denissenkov et al. 2014) is classical novae. José et al. (2004) pointed out that the main effect of classical nova nucleosynthesis on the Si isotopes is an increase of $\delta^{30}$Si/$^{28}$Si with $\delta^{29}$Si/$^{28}$Si below or approaching zero. Putative nova, C1, and C2 grains are compared to ONe nova model predictions of José et al. (2004) with updated nuclear inputs for Si isotope ratios in Fig. 3. The nova model predictions are shown as mixing lines between weighted average isotopic compositions of pure ONe nova ejecta and the solar isotopic composition. This figure shows that all the putative nova grains lie between the 1.15 and 1.35 $M_\odot$ ONe nova model predictions, although the pure nova ejecta need to be diluted with a large amount of isotopically normal material to match the grain data. In contrast, type C2 grains show enrichments in both $^{29}$Si and $^{30}$Si relative to $^{28}$Si that are incompatible with the nova nucleosynthetic calculations. For instance, even though individual shells of nova ejecta are more variable with respect to their bulk compositions, the lowest ratio of $\delta^{30}$Si/$^{28}$Si to $\delta^{29}$Si/$^{28}$Si reached in individual shells of the 1.35 $M_\odot$ ONe nova model (~6) is still a factor of three to six higher than those of C2 grains (Fig. 5 of Amari et al. 2001b), which cannot be explained by mixing with materials of solar isotopic composition (i.e., isotopic dilution cannot cause the ratio of $\delta^{30}$Si/$^{28}$Si to $\delta^{29}$Si/$^{28}$Si to vary). Thus, ONe nova models are incompatible with the Si isotopic signatures of C2 grains.

Other types of stars that can produce low $^{12}$C/$^{13}$C ratios include born-again AGB stars and J-type stars. Born-again AGB stars (Fujimoto 1977) are post-AGB stars that undergo a very late thermal pulse (VLTP, e.g., Sakurai's object). At this stage, the remnant star has only a thin residual H shell left that could be ingested into the convective He-rich shell during the VLTP, producing extremely low $^{12}$C/$^{13}$C ratios. However, models of such stars predict large excesses in $^{14}$N (Fig.10 of Herwig et al. 2011) and are thus unlikely as sources of the $^{13}$C- and $^{15}$N-rich SiC grains. J-type giants are C-rich stars that are mainly identified by their low $^{12}$C/$^{13}$C and the absence of *s*-element enrichments (Abia & Isern 2000); their origin is poorly understood. Hedrosa et al. (2013) reported N isotope ratios for six J-type giants, one of which, WX cyg, may have a $^{14}$N/$^{15}$N ratio as low as six and could show the same $^{15}$N-excess seen in the putative nova grains. Thus, C-rich J-type stars might have contributed $^{13}$C- and $^{15}$N-enriched SiC dust grains to the Solar System. However, according to the $^{29,30}$Si excesses in C2 grains (see also Section 4.1.2 for details), *s*-process element enrichments are expected in these grains, while such enrichments are not observed in J-type stars (Abia & Isern 2000). Thus, at present they are less likely to be the stellar progenitors of C2 grains.

*4.1.2 Excesses in $^{29}$Si and $^{30}$Si of Type C2 Grains: Neutron Burst Signature in CCSNe*

Similar to type C2 grains, another rare group of presolar SiC grains, type C grains (hereafter type C1 grains), also have large excesses in $^{29}$Si and $^{30}$Si, relative to $^{28}$Si. However, C1 grains are enriched in $^{12}$C and $^{15}$N, relative to solar materials, similar to the isotopic signatures of most X grains (~1−2% of presolar SiC grains, see Zinner 2014 for references). Figure 3 shows that the Si isotopic compositions of C1 and C2 grains fall on similar trends. For reference, the slopes of the linear fits for C1 and C2 grains in Fig. 3 are 0.89±0.05 and 1.12±0.57, respectively, which are consistent with each other within 95% confidence bands. In contrast to the $^{28}$Si-enriched X grains that came from CCSNe, the rarer type C1 and C2 grains (< 0.1%) are characterized by large excesses in $^{29}$Si and $^{30}$Si. These are most likely explained by neutron capture triggered by activation of the $^{22}$Ne($\alpha,n$)$^{25}$Mg neutron source during CCSN explosions (Meyer et al. 2000; Pignatari et al. 2013a). The associated neutron-capture nucleosynthesis is called a neutron burst process, *i.e.*, *n* process (Blake & Schramm 1976; Thielemann et al. 1979; Meyer et al. 2000). Figure 4 shows the 12 $M_\odot$, $Z_\odot$ CCSN model of Woosley & Heger (2007, W&H07), which predicts a neutron burst zone in the outer C/O and inner He/C zones (shaded area in the left panel of Fig. 4). Model predictions for the Si isotope ratios of these shells are compared with those of C1 and C2 grains (right panel of Fig. 4); all the grain data are well matched by mixing neutron-burst products with H envelope material (with an assumed initial solar Si isotopic composition). The good agreement is consistent with the $^{29}$Si and $^{30}$Si excesses of both C1 and C2 grains being the result of neutron-burst nucleosynthesis in CCSNe.

Pignatari et al. (2013a) proposed that the large negative $\delta^{33}$S/$^{32}$S and $\delta^{34}$S/$^{32}$S values found in some C1 grains can be explained by the presence of $^{32}$S from the in situ decay of short-lived $^{32}$Si ($\tau_{1/2}$ = 153 a) produced by the *n* process (predicted $^{32}$Si/$^{28}$Si on the order of $10^{-3}$, Fig. 5 of Pignatari et al. 2013a), which could be incorporated into SiC grains as $^{32}$Si and then decay to $^{32}$S after grain condensation. This 12 $M_\odot$ CCSN model, as discussed in Xu et al. (2015), predicts high $^{32}$Si/$^{28}$Si ratios in the neutron burst zone (Fig. 4), in agreement with the 15 $M_\odot$ model by Pignatari et al. (2013b). The Si and S isotopic compositions of C1 grains therefore can be explained by the 12 $M_\odot$ CCSN model predictions for the C-rich neutron burst zone, if mixing with H envelope material is considered.

In addition to the good agreement between type C2 grain data and neutron burst predictions for Si isotopes, additional neutron-burst isotopic signatures were observed in the grain GAB and also strongly support the CCSN origin (Fig. 5). To qualitatively compare model predictions with the grain data, we mixed materials from individual neutron burst shells of different CCSNe with different amounts of H envelope material (with solar Si and Ti isotopic compositions to dilute the *n*-process products) to match the measured $\delta^{29}Si/^{28}Si$ value of GAB (230±6‰) and calculated predictions for the other neutron-rich Si and Ti isotopes. Figure 5 clearly shows that all of the baseline CCSN models (*e.g.*, 12 and 25 $M_\odot$ CCSN model by W&H07) predict elevated $\delta^{30}Si/^{28}Si$, $^{32}Si/^{28}Si$, and $\delta^{50}Ti/^{48}Ti$ values in the neutron burst zone, and the 25 $M_\odot$ CCSN model predictions agree best with the grain data. Note that the 25 $M_\odot$ CCSN model of W&H07 predicts that the neutron burst zone is located within the C/O zone, where the C/O ratio is **less** than unity. So the main prerequisite for SiC formation (C>O) therefore cannot be satisfied. Overall, the Si, S, and Ti isotopic compositions of grain GAB are well reproduced by pure signatures of the neutron burst zone without the need of further mixing from deeper zones. In addition, no measurable excess in $\delta^{44}Ca/^{40}Ca$ was found in grain GAB, indicating that its parental material was mainly from outer zones of the supernova without significant incorporation of $^{44}Ti$ from deeper layers, in good agreement with the constraints from other *n*-process isotopic signatures.

CCSN models, however, predict extremely high $^{12}C/^{13}C$ ratios for the C/O and He/C zones ($10^7$–$10^9$ for $^{12}C/^{13}C$, Table 2 of Xu et al. 2015 and references therein), because $^{12}C$ is abundantly produced as a result of the triple-alpha reaction, whilst the lack of H in these zones precludes abundant production of $^{13}C$. Xu et al. (2015) therefore highlighted difficulties in explaining $^{13}C$ and $^{15}N$ excesses in SiC X grains and low density graphite grains that were discussed in previous studies (e.g., Travaglio et al. 1999, Yoshida 2007). Nittler & Hoppe (2005) pointed out that the CCSN isotopic signatures (excesses in $^{28}Si$, $^{44}Ca$, and $^{49}Ti$) observed in the $^{13}C$- and $^{15}N$-enriched grain M11-334-2 demonstrates that high-temperature H burning must occur in the He-shell zones of at least some CCSNe and they proposed that some kind of extra mixing process(es) brought extra H into the He shell of a pre-supernova star, which could eventually enhance $^{13}C$ and $^{15}N$ production in the He shell via proton-capture reactions at high temperatures. For the first time, CCSN models of P15 show that an H ingestion event in the pre-supernova phase can affect the usual He-burning signature, and the impact of the following explosive H burning in the He shell during the supernova shock is considered in the models. In the next Section, we compare our new grain measurements on C, N, and Al isotopic abundances with theoretical predictions by P15, and we analyze their main nucleosynthetic features.

*4.1.3 $^{13}$C and $^{15}$N Excesses of Type C2 Grains: Proton Capture Signatures in CCSNe*

In addition to extra mixing processes occurring along the boundary of two convective zones caused by e.g., rotation (e.g., Meynet et al. 2006) and gravity waves (Arnett & Meakin 2011), more tumultuous mixing events can occur in stars, such as shell mergers (e.g., Woosley & Weaver 1995; Rauscher et al. 2002), and convective-reactive events following the ingestion of less processed material into hotter and deeper stellar zones (e.g., Herwig et al. 2014). For instance, Woosley & Weaver (1995) found that in one finely zoned 10 $M_\odot$ model of solar metallicity, the He shell is fully convectively linked with the H envelope, which could possibly bring a large amount of H into the He/C shell in CCSNe. Similarly, Pignatari et al. (2013b) found that H is ingested into the He shell in a 25 $M_\odot$ supernova model with solar metallicity.

Recently, P15 have qualitatively investigated the effect of H ingestion in the He shell material (e.g., Herwig et al. 2011; Stancliffe et al. 2011; Herwig et al. 2014; Woodward et al. 2015) and the presence of residual H during explosive He burning in the 25 $M_\odot$ supernova model of Pignatari et al. (2013b). In particular, during pre-supernova evolution, H was ingested in the He shell and shell convection was then turned off, allowing survival of the ingested H in the He shell. P15 adopted two sets of supernova shock models ($T_{peak}$ = 0.7×10$^9$ K at the bottom of the He-rich zone in model 25d, and $T_{peak}$ = 2.3×10$^9$ K in model 25T, with the latter reproducing the stellar condition of the 15$M_\odot$ CCSN model 15d in Pignatari et al. 2013b). The amount of H remaining from the pre-supernova evolution is varied from 1.2%(-H, the value predicted by the stellar model 25d), 0.12%(-H10), to 0.0024% (-H500) in the He shell. The predictions of these models for the He/C zone are compared to the $^{13}$C- and $^{15}$N-enriched presolar SiC grains from this study in Fig. 6. Comparing the predictions of the -H500 (the least amount of H ingested) models with those of the other two models clearly indicates that explosive H burning greatly enhances the production of $^{13}$C, $^{15}$N, and $^{26}$Al in the He/C zone.

Three of the four C2 grains have $^{12}$C/$^{13}$C ratios lower than three, which can only be reached by the 25T-H and 25d-H models. Although Al contamination could lower the initial

$^{26}$Al/$^{27}$Al ratios estimated for the grains, the inferred amount of Al contamination found in most SiC grains is only up to 65%, i.e., a factor of two higher at most (Groopman et al. 2015), which generally agree with the amount of Al contamination found in the grain GAB, $20^{+60}_{-15}$ % (Fig. 2b). By mixing $^{13}$C- and $^{15}$N-enriched shells with H envelope material to match C isotope ratios of C2 grains, we found that the 25d-H predictions are more comparable ($^{26}$Al/$^{27}$Al≈0.02) to the grain data ($^{26}$Al/$^{27}$Al≈0.01−0.03) than the 25T-H model predictions ($^{26}$Al/$^{27}$Al≈0.1−0.3). Finally, we point out that as indicated by their $^{28}$Si and $^{44}$Ti excesses, X grains are likely to have incorporated more material from deeper layers, e.g., from the C/Si zone (Pignatari et al. 2013b). So even if explosive H burning also occurred in the outer He shells of their parent CCSNe, the isotopic signatures of explosive H burning can be modified by incorporating materials with $^{12}$C-rich explosive He-burning signatures in X grains.

*4.1.4 Constraints on the Neutron Burst Environment during Explosive H Burning*

Figure 5 shows differences in the predicted Si and Ti isotopes of the CCSN models for the neutron burst zone. In the 25d-H to 25d-H500 models, no resolvable neutron burst zone can be found and these are therefore not shown. A neutron burst zone exists in the 25T-H500 model, but there is a limited amount of H ingestion and the effect of proton-capture nucleosynthesis is negligible (Fig. 5). Differences are found between P15 and W&H07 model predictions in Fig. 5. For instance, while the 25d-H500 model predictions for $^{32}$Si/$^{28}$Si and δ$^{50}$Ti/$^{48}$Ti lie between the 12 and 25 $M_\odot$ CCSN model predictions of W&H07, the corresponding predictions for δ$^{30}$Si/$^{28}$Si are higher than both the W&H07 models. Together with the intrinsic differences between the two sets of models, the variations that we consider here could be caused by different nuclear cross sections adopted for the nucleosynthetic calculations of W&H07 and P15. For neutron capture on $^{30}$Si, W&H07 adopted the experimental MACS (Maxwellian Averaged Cross Section) at 30 keV from Bao et al. (2000), 6.5±0.6 mb. On the other hand, P15 adopted the more recent $^{30}$Si MACS values reported by Guber et al. (2003), which is more than a factor of three times lower (1.82±0.33 mb) than that used by W&H07 at 30 keV. Because of the lowered probability for neutron capture on $^{30}$Si by adopting the Guber et al. (2003) MACS values, the P15 models predict higher $^{30}$Si abundance relative to that of $^{28}$Si. Thus, the difference in the δ$^{30}$Si/$^{28}$Si predictions is mainly caused by different $^{30}$Si MACS values adopted in the two sets of models.

When the ingested amount of H is increased from 0.0024% to 1.2% (25T-H model), the neutron burst nucleosynthesis is suppressed because the $^{22}$Ne neutron source abundance ($^{22}$Ne($\alpha$,n)$^{25}$Mg) for the *n* process is greatly reduced due to competing proton captures on $^{22}$Ne via $^{22}$Ne(p,$\gamma$)$^{23}$Na (P15). Consequently, $^{49}$Ti is more abundantly made than $^{50}$Ti, and $^{32}$Si/$^{28}$Si is about 20 orders of magnitude lower, because of the lowered neutron density and the consequent negligible production of $^{32}$Si (Fig. 5). Nucleosynthetic calculations based on one-dimensional (1D) stellar models by P15 therefore cannot explain the *n*-process isotopic signatures of the C2 grains. Highly likely, the discrepancy between the grain data and theoretical predictions is caused by the fact that 1D stellar models cannot obviously predict the large heterogeneities of H ingestion into the He shell that have been found in multi-dimensional hydrodynamic simulations (e.g., Herwig et al. 2014). Thus, although 1D nucleosynthetic calculations can provide robust predictions for specific nuclide abundances **resulting** from the H ingestion event when the relevant reaction rates are experimentally well determined (e.g., reactions to produce $^{12}$C, $^{13}$C, $^{14}$N, $^{15}$N, $^{26}$Al, $^{27}$Al), different neutron- and proton-capture isotopic signatures may exist in different regions of the He shell, resulting from the multi-dimensional nature of H-ingestion events, i.e., varying amounts of H mixed into different regions of the He shell due to the heterogeneous H-ingestion process predicted by multi-dimensional models.

Overall, the P15 models suggest that H has to be effectively injected into the He shell (~1.2%) in order to account for the low $^{12}$C/$^{13}$C isotope ratios of C2 grains. Furthermore, the neutron burst isotopic signatures of C2 grains strongly support the heterogeneous distribution of ingested H and asymmetries by multi-dimensional simulations for the H-ingestion event. That is to say, C2 grains probably incorporated materials from different regions of the He shell, so proton-capture isotopic signatures are the result of explosive H burning in regions with higher amounts of ingested H (~1.2%), while *n*-process isotopic signatures are records of nucleosynthesis during pre-supernova evolution in regions with negligible amounts of ingested H (<0.0024%). Therefore, although 1D stellar models provide important insights into different nucleosynthetic processes in CCSNe, complete multi-dimensional hydrodynamic simulations of the H ingestion event are needed to illustrate how parental materials of C2 grains are mixed prior to their condensation and to quantitatively compare nucleosynthetic predictions with C2 grain data. Currently, detailed multi-dimensional simulations for H ingestion are only available for the He shell in low mass stars. Therefore, similar calculations are also needed for massive stars with solar-like metallicity. **To conclude, comparison of the C2 grain data with nucleosynthetic calculations based on 1D stellar models is critical to capture physical process(es) that are not previously considered, while the multi-element isotope data of C2 grains also presents a challenge for state-of-the-art 1D simulations of the hydrogen ingestion event, which provides a unique opportunity to constrain both neutron- and proton-capture environments prior to and during the supernova explosions for multi-dimensional hydrodynamic simulations.**

4.2 Parent Stars of Putative Nova Grains: Novae or Supernovae?

Hydrogen burning through the CNO cycle at high temperatures can yield both low $^{12}C/^{13}C$ and $^{14}N/^{15}N$ ratios (Section 4.1.1), and could occur in both novae and supernovae. During classical nova outbursts, the main nucleosynthetic path is driven by ($p,\gamma$), ($p,\alpha$), and $\beta^+$ processes, with little contribution from $n$- and $\alpha$-capture processes (Fig. 2 of José et al. 2012). In contrast, all of these processes could occur in the He/C zone during CCSN explosions, assuming ingestion of H depending on the peak temperature, density, and the amount of ingested H.

We determined Ca-Ti isotope ratios in four of the seven putative nova grains from this study, all of which have solar $^{44}Ca/^{40}Ca$ and Ti isotope ratios. Although these isotopic signatures are in good agreement with model predictions for classical nova nucleosynthesis, they do not rule out explosive H burning during supernova explosions, because the normal isotopic compositions could be simply caused by surface contamination, as in the case of Al (see Section 3.2). Even if we assume negligible amounts of surface contamination, the normal Ca-Ti isotope ratios are still consistent with CCSN nucleosynthesis because (1) $^{44}Ti$ is not produced in the He shell. So $^{44}Ca$ excesses from $^{44}Ti$ decay would not be expected in grains that condensed there, in agreement with the solar $^{44}Ca/^{40}Ca$ ratio of putative nova grains; (2) additionally, Ti isotope anomalies are predicted to be only a factor of two higher than those of Si in the outer He shell, and this corresponds to predicted Ti isotope anomalies lower than 200‰ in these four grains, in agreement with the grain data within uncertainties. Nucleosynthetic processes in different CCSN zones are quite diverse and the resulting isotopic signatures become complicated by including H ingestion and *ad hoc* mixing among different zones. As a result, proton-capture isotopic signatures of classical nova nucleosynthesis could be produced by explosive H burning in CCSNe, with contribution from nova-like isotopic signatures of the O zone in the 25T-H model of P15. Although Parikh et al. (2014) have recently pointed out enhancement in the $^{33}$S/$^{32}$S ratio as a diagnostic isotopic signature of nova grains, in fact, the 25T-H model predictions for the $^{33}$S/$^{32}$S ratio are also positive in both the O (nova-like) zone and the He/C zone of CCSNe. Thus, the $^{33}$S/$^{32}$S ratio is not specifically diagnostic of grains from novae, but more in general of H-burning at high temperatures in stars.

In the following Sections, we first examine if classical nova models can explain the grain data. If not, the possibility of explaining these isotopic signatures by CCSNe will be discussed. We remind the readers that C-rich J-type stars also remain a possibility as the stellar site of putative nova grains. However, due to the lack of knowledge about the origin and detailed nature of these stars, nothing is known about isotopic signatures of elements other than C and N (astrophysical observations, e.g., Abia & Isern 2000) in them. In addition, recent observations of Nova Vul 1670 showed that the object is best explained as the remnant of a merger of two stars that were surrounded by an outflow in the form of O-rich molecular gas that was extremely enriched in $^{13}$C and $^{15}$N ($^{12}$C/$^{13}$C~2, $^{14}$N/$^{15}$N~3, Kamiński et al. 2015). Although events exactly like Vul 1670 are unlikely to be the progenitors of putative nova SiC grains because of its oxidized environment, it remains an open question if there exist C-rich Vul 1670 like objects that are enriched in $^{13}$C and $^{15}$N that could be the progenitors of putative nova grains.

*4.2.1 Putative Nova Grains versus Nova Nucleosynthesis*

A comparison of putative nova grains and nova models was given by Amari et al. (2001b). The new putative nova grains in our study span the same range as previous grains for C, N, and Si isotopic compositions (Fig. 1), so the reader is referred to Amari et al. (2001b) for discussion on these isotope ratios. In this section, we focus on comparison of the grain data with nova model predictions for $^{26}$Al/$^{27}$Al and S isotope ratios.

Figure 7(a) shows that mixing curves for ONe and CO nova model predictions follow different trajectories in the plot of $^{26}$Al/$^{27}$Al versus $^{12}$C/$^{13}$C, mainly because $^{26}$Al is highly overproduced in ONe novae, which occur at higher outburst temperatures. Overall, four putative nova grains with $^{26}$Al/$^{27}$Al >0.06 can be roughly matched by ONe nova model predictions. The mixing ratios required to match the grain data indicate that if they are indeed ONe nova grains, they must have come from ONe novae with masses lower than 1.35 $M_\odot$, because the extreme $\delta^{30}Si/^{28}Si$ values predicted by the 1.35 $M_\odot$ ONe nova model are not observed (Fig. 3). In addition, putative nova grains with $^{26}Al/^{27}Al$ ~0.01 overlap with CO nova model predictions, in contrast to the previous argument that all of the putative nova grains originated from ONe novae solely based on Si isotope ratios (Amari et al. 2001b). The Si isotope anomalies of these CO nova grains are within ±100‰ with small excesses in $^{30}Si$ relative to $^{29}Si$, while the main effect of CO nova nucleosynthesis is to decrease $\delta^{29}Si/^{28}Si$ while leaving $\delta^{30}Si/^{28}Si$ unchanged. Variations in the Si isotopes of mainstream presolar SiC grains (±200‰) are believed to result from the effect of GCE on their parent stars (Alexander & Nittler 1999; Lugaro et al. 1999; Zinner et al. 2006). Thus, Si isotope ratios of the companion main-sequence star in a binary system that forms a nova outburst latter should fall along a line with a slope of 1.38 defined by the mainstream grain data in the plot of $\delta^{29}Si/^{28}Si$ versus $\delta^{30}Si/^{28}Si$ (Zinner et al. 2007). Consequently, CO nucleosynthesis could lower the $\delta^{29}Si/^{28}Si$ value during nova outburst and therefore result in excess $^{30}Si$ relative to $^{29}Si$ as observed in putative CO nova grains. As a result, the possibility of grains from CO novae mitigates one of the difficulties in linking putative nova grains to novae (problem 3 discussed in Section 1), but thermodynamic calculations for the innermost shell of CO novae predict that SiC cannot be condensed in such oxidized environment (José et al. 2004). Condensation calculations including kinetic effects and mixing among different shells of CO novae, however, are needed to fully investigate this issue.

More importantly, none of the nova models can explain the $^{34}S$ anomalies found in two putative nova grains (G270_2 & Ag2_6). As shown in Fig. 1 of José et al. (2012), the proton-capture path cannot reach the S mass region in novae with white dwarf masses lower than 1.35 $M_\odot$, so that S isotope abundances are barely affected by lower-mass nova nucleosynthesis. In 1.35 $M_\odot$ ONe novae, the proton-capture nucleosynthesis overproduces S isotopes, and $^{32,33}S$ isotopes are more abundantly made than $^{34}S$, corresponding to negative $\delta^{34}S/^{32}S$ values. The 1.35 $M_\odot$ ONe nova model, however, predicts huge excesses in $\delta^{30}Si/^{28}Si$, inconsistent with the grain data (Fig. 7b). In addition, neutron-rich isotopes cannot be made in novae because nova nucleosynthesis is dominated by proton-capture and $\beta^+$ decay (Fig. 2 of José et al. 2012). Negative $\delta^{34}S/^{32}S$ values of nova grains therefore cannot be explained by $^{32}S$ excess from $^{32}Si$ $\beta^-$ decay as argued for the C1 and C2 grains from CCSNe. Thus, the isotopic compositions of grains G270_2 and Ag2_6 seem to be inconsistent with predictions of classical nova nucleosynthesis.

However, it is noteworthy to point out that the production of S isotopic abundances is still affected by nuclear uncertainties, e.g., the treatment of the $^{34}$Cl ground and isomeric nuclear states (e.g., Coc et al. 2000). Furthermore, these nova nucleosynthetic models are all based on 1D hydrodynamic model calculations that cannot predict asymmetries during nova outbursts and do not account for the effect of rotation. Thus, the mismatch with the multi-element isotopic data by 1D nova model predictions could also indicate incorporation of products from regions with varying nucleosynthetic environments within the nova ejecta during a single nova outburst (similar to the discussion for CCSNe in Section 4.1.4). Thus, multi-dimensional hydrodynamic calculations of nova outbursts (e.g., Casanova et al. 2011) are needed to confirm the mismatch of nova model predictions with the isotopic compositions of G270_2 and Ag2_6.

*4.2.2 Putative Nova Grains versus CCSN Nucleosynthesis*

In Fig. 6, predictions for explosive H burning events in CCSNe are overlapped with most of the putative nova grain data, if one considers variations in the peak temperature and peak density during explosions, the amount of ingested H, and the percentage of mixed H envelope material. This demonstrates the strong similarities between explosive H burning events in novae and CCSNe. Thus, many isotopic signatures can be diagnostic for both stellar sources. The explosive H ingestion scenario in CCSNe, however, at the moment has many advantages over classical novae in explaining the putative nova grain data: (1) Explosive H burning in CCSNe only affects a thin shell of the He/C zone, and mixing with a large amount of isotopically close-to normal material should be a natural consequence of grain condensation during explosions. In contrast, a mechanism for mixing such a large amount of normal material with pure nova ejecta has not been identified; (2) A C-rich environment is obtainable by mixing explosive H ingestion products with extended He-shell layers and more external H-rich material. The main prerequisite for SiC formation (C>O) can therefore be satisfied.

While S isotopic anomalies cannot be explained by models for low-mass novae (<1.35 $M_\odot$), the negative $\delta^{34}S/^{32}S$ values of grains G270_2 and Ag2_6 could be matched by local mixing between the He/C zone and zones with $^{28}$Si, $^{32}$S excesses in CCSNe (Fig. 8). Traditionally, the Si/S zone of the classical W&H07 models was considered as the only zone with large excesses in $^{28}$Si and $^{32}$S because of alpha captures. Pignatari et al. (2013b) have recently shown that increased shock velocities and consequently higher peak temperatures can result in efficient α-capture at the bottom of the He/C zone, and a C/Si zone can be formed there. The 25T-H model predicts such a C/Si zone (blue circles in Fig. 8), which lies below the $^{13}$C- and $^{15}$N-enriched shells. As shown in Fig. 8, local mixing between the He/C and C/Si zones can explain both the $^{34}$S depletion (or $^{32}$S excess) and $^{30}$Si excesses of G270_2 and Ag2_6. Thus, the state-of-the-art model calculations suggest that grains G270_2 and Ag2_6 might have originated from CCSNe. Negligible amounts of $^{32}$Si are produced in the He/C zone (Fig. 5) and the C/Si zone of the 25T-H model, so the negative $\delta^{34}S/^{32}S$ values cannot be explained by $^{32}$S excesses from radioactive $^{32}$Si decay in this scenario.

Finally, the stellar origin of $^{15}$N-enriched AB grains is even more ambiguous than that of the putative nova grains. J-type stars, CCSNe with H-ingestion, and CO novae are all possible stellar sites, whilst the CO novae are less favored because of the oxidized stellar environment, which make it less likely for SiC to condense. It should also be noted that there are consistent differences between the $^{15}$N-enriched AB grains and putative nova grains (higher $^{14}$N/$^{15}$N, lower $^{26}$Al/$^{27}$Al, no or smaller excesses in $^{30}$Si relative to $^{29}$Si in AB grains, see Section 3.1 for details). This seems to indicate that $^{15}$N-enriched AB grains are carriers of weaker proton-capture processes compared to putative nova grains.

### 4.3 Additional Evidence for Presolar Dust Grains from Novae?

Novae can produce significant amounts of $^{22}$Na with a half lifetime of 2.6 yr, which could be incorporated into condensing SiC grains where it would decay *in situ* to $^{22}$Ne (Starrfield et al. 1997). Heck et al. (2007) measured isotopic compositions of Ne in single presolar SiC grains and identified one out of the 110 grains analyzed as an AB grain with excess in $^{22}$Ne, which was proposed as an indication of CO nova origin. In fact, CCSN models also predict significant production of $^{22}$Na. For instance, the CCSN models of W&H07 predict that the $^{22}$Ne/$^{20}$Ne ratio can reach ~10 in the He/C zone, which could explain the grain data ($^{22}$Ne/$^{20}$Ne < 0.36) by mixing with isotopically normal material (solar $^{22}$Ne/$^{20}$Ne=0.1). Additionally, P15 models predict large amounts of $^{22}$Na produced in the He shell that could be incorporated into C-rich dust grains and *in situ* decay to $^{22}$Ne. Thus, both CCSNe and novae can produce higher-than-solar $^{22}$Ne/$^{20}$Ne isotopes, which are therefore not diagnostic of a nova origin. To conclude, none of the isotopic signatures that have been found in putative nova SiC grains so far unequivocally link the grains to a nova origin.

A few highly $^{17}$O-enriched presolar silicates and oxides have been found in primitive meteorites in previous studies (Gyngard et al. 2011; Leitner et al. 2012; Nguyen & Messenger 2014), which might have originated from novae. These grains generally show enrichments in $^{25}$Mg and $^{26}$Mg (the latter ascribed to $^{26}$Al decay), in agreement with both supernova and nova model predictions. In addition, large excess in $^{30}$Si relative to $^{29}$Si was found in one of the two silicate grains studied by Nguyen & Messenger (2014), which is similar to the Si isotopic signatures of putative nova SiC grains. The highest $^{17}$O/$^{16}$O ratio predicted by the H-ingestion models of P15 is only up to 2×10$^{-3}$ (25T-H model), which is about one order of magnitude lower than the highest ratio observed in the $^{17}$O-enriched silicate and oxide grains. Thus, the O isotopic compositions of the putative nova silicates and oxides are in better agreement with CO nova nucleosynthetic model predictions of José et al. (2012) than are the putative nova SiC grains.

## 5. CONCLUSIONS

Multi-element isotope data clearly show that the rare population of presolar SiC grains with very large enrichments in both $^{13}$C and $^{15}$N, previously argued to mainly originate in classical novae, can be divided into two groups, referred to here as C2 grains and putative nova grains. Based on detailed comparisons of new and literature isotope data on these grains with state-of-the-art nucleosynthetic calculations, we can come to the following conclusions:

(1) Type C2 grains are characterized by large $^{13}$C and $^{15}$N enrichments, and $^{29}$Si and $^{30}$Si excesses, which incorporated materials from regions of CCSNe similar to the $^{29}$Si- and $^{30}$Si-enriched C1 grains, but with higher $^{12}$C/$^{13}$C ratios (>10). The *n*-process isotopic signatures of C2 grains (excesses in $^{29,30}$Si, $^{50}$Ti) are well explained by neutron-burst nucleosynthesis in CCSNe, pointing towards CCSN origin. The $^{13}$C and $^{15}$N enrichments of C2 grains therefore strongly suggest the occurrence of explosive H burning in the He shell during CCSN explosions. By ingesting ~1.2% H into the He shell, the P15 models predict that the neutron source for the *n* process, $^{22}$Ne, is consumed by proton capture during explosive H burning, so that *n* process cannot occur. In contrast to the 1D model predictions, type C2 grains show both proton- and neutron-capture isotopic signatures with the *n*-process signatures unaffected by the proton capture process, thus strongly supporting the large heterogeneity predicted for H ingestion into the He shell by multi-dimensional simulations. **Therefore, while comparison with 1D stellar models helps to capture missing physical processes in stellar models, complete multi-dimensional model calculations for hydrogen ingestion events in massive stars at solar and close-to-solar metallicity are needed to match the multi-element isotope data.** Presolar grain data allows derivation of stringent constraints on both the neutron- and proton-capture environments in the He shell of CCSNe in detail. We point out that because of the uncertain physical mechanism(s) responsible for mixing H into the He shell in pre-supernovae, there still remain a number of open questions about H-ingestion events, e.g., what fraction of CCSNe experience explosive H burning? How varying is the effect of explosive H burning on nucleosynthetic products of the He/C zone in different CCSNe?

(2) Putative nova grains are characterized by $^{13}$C and $^{15}$N enrichments and large excesses in $^{30}$Si relative to $^{29}$Si. In fact, nucleosynthetic model predictions of explosive H burning events in novae and CCSNe are quite similar for a number of elements, so the stellar origin of putative nova grains is quite ambiguous. Comparison with nova models shows that if putative nova grains are from novae, (1) they came from both CO and ONe novae; and (2) the mass of their parent stars should be lower than 1.35 $M_{\odot}$. The current nova models, however, cannot explain the negative $\delta^{34}$S/$^{32}$S ratios found in two putative nova grains from this study. Instead, local mixing in He-shell material of CCSNe can reproduce these isotopic signatures. It is therefore more likely that these two putative nova grains were actually sourced from CCSNe. Putative nova oxides and silicates, however, show extremely high $^{17}$O/$^{16}$O ratios, while H-ingestion model predictions of P15 are an order of magnitude lower and therefore cannot explain the grain data. So far, $^{17}$O-enriched oxides and silicates seem to be the most likely nova grain candidates.

(3) Nitrogen-15 enriched AB grains show systematic differences compared to putative nova grains, such as higher $^{14}$N/$^{15}$N ratios (>70), and lower $^{26}$Al/$^{27}$Al ratios. Also, they are overlapped with mainstream grains for Si isotope ratios, and lack the $^{30}$Si excesses found in putative nova grains. J-type stars, CO novae, and CCSNe are all potential stellar progenitors of $^{15}$N-enriched AB grains. The distinctive differences in N, Si, and Al isotope ratios between AB and nova grains indicate either that they came from different types of stars, or that they came from the same stellar source, but incorporated stellar materials processed in different nucleosynthetic conditions.

Acknowledgements: We thank Alexander Heger for providing CCSN nucleosynthetic model calculations. This work was supported by NASA's Cosmochemistry program (grant NNX10AI63G to LRN). MP acknowledges significant support to NuGrid from NSF grants PHY 09-22648 (Joint Institute for Nuclear Astrophysics, JINA), NSF grant PHY-1430152 (JINA Center for the Evolution of the Elements) and EU MIRGCT-2006-046520. MP acknowledges the support from the "Lendulet-2014" Program of the Hungarian Academy of Sciences (Hungary) and from SNF (Switzerland). JJ acknowledges the Spanish MINECO grants AYA2013-42762-P and AYA2014-59084-P, and the Generalitat de Catalunya grant SGR0038/2014 for support.

FIGURE CAPTIONS

Figure 1 Isotopic compositions of C, N, Si, and Al of three new $^{15}$N-enriched AB grains, four C2 grains, and seven new putative nova grains from this study, in comparison to 57 $^{15}$N-enriched AB grains (Hoppe et al. 1994, 1996; Amari et al. 2001a; Huss et al. 1997), and eight putative nova grains (Gao & Nittler 1997; Amari et al. 2001b; Nittler & Hoppe 2005) reported in the literature. The isotope data are plotted with 1σ uncertainties. The dashed lines represent terrestrial isotopic compositions. For $^{14}$N/$^{15}$N, the value of protosolar nebula (441±5) reported by Marty et al. (2011) is also shown for reference. Data sources of C1 grains: Pellin et al. (2000); Croat & Stadermann (2008), Gyngard et al. (2010), Hoppe et al. (2012), and Xu et al. (2015).

Figure 2 NanoSIMS depth profile of grain GAB (type C2, see text for details) is shown in (a). During the first five cycles of the profile, the Si and Mg signals are increasing at different rates, while the Al signal is decreasing, possibly because of surface Al contamination. This results in deviation of the grey data points during the first five cycles from the linear correlation defined by the rest of the data points after equilibration of the ion signals in (b). The weighted ODR (Orthogonal Distance Regression) method is a linear least-squares fitting that minimizes scatter orthogonal to the best fit line and considers uncertainties in both x-axis and y-axis for each data point. The isotope data are plotted with 1σ uncertainties.

Figure 3 Plot of δ$^{29}$Si/$^{28}$Si versus δ$^{30}$Si/$^{28}$Si. Putative nova, and types C1 and C2 grains are plotted in comparison to ONe nova model predictions shown as solid black lines. The number next to each line indicates the initial mass of the ONe white dwarf. The grey

areas highlight Si isotopic compositions produced by the *n*-process. The isotope data are plotted with 1σ uncertainties.

Figure 4 Left panel: plot of 12 $M_\odot$ CCSN model predictions for $^{29}Si/^{28}Si$, $^{30}Si/^{28}Si$, $^{49}Ti/^{48}Ti$, $^{50}Ti/^{48}Ti$, and $^{32}Si/^{28}Si$ in the mass coordinate. Note that isotope ratios except $^{32}Si/^{28}Si$ are normalized to solar values (indicated by the legend), and unity corresponds to the solar isotopic composition. The shaded area is the neutron burst zone, where the *n*-process takes place. Right panel: in the three isotope plot of $\delta^{29}Si/^{28}Si$ versus $\delta^{30}Si/^{28}Si$, C1 and C2 grains are compared to 12 $M_\odot$ CCSN model predictions for the shaded zones in the left panel. The isotope data are plotted with 1σ uncertainties. The dashed lines are mixing lines between two end-members with extreme Si isotopic compositions and H envelope material. The shaded area therefore represents isotopic compositions that can be explained by the model predictions.

Figure 5 Si- and Ti-isotopic compositions of grain GAB are compared to different CCSN model predictions for the neutron burst zone, which are normalized to $\delta^{29}Si/^{28}Si$ at 230‰ by assuming different dilution factors for different shells. The isotope data are plotted with 1σ uncertainties. CCSN models include (1) 12 and 25 $M_\odot$ CCSN model predictions by Woosley & Heger (2007, W&H07); and (2) 25 $M_\odot$ CCSN model predictions with 1.2% (25T) and 0.0024% (25T-H500) of H ingested into the He shells by Pignatari et al. (2015, P15).

Figure 6 Isotopic compositions of C, N, Si, and Al of three new $^{15}N$-enriched AB grains, four C2 grains and seven new putative nova grains from this study are compared to H-ingestion CCSN model predictions for the He/C zone by P15. "Inner" refers to the innermost shell of the He/C zone and "Outer" refers to the outermost shell. The isotope data are plotted with 1σ uncertainties. The dashed lines represent terrestrial isotopic compositions. For $^{14}N/^{15}N$, the solar value reported by Marty et al. (2011) is also shown.

Figure 7 Plots of $^{26}Al/^{27}Al$ versus $^{12}C/^{13}C$, and $\delta^{34}S/^{32}S$ versus $\delta^{30}Si/^{28}Si$. Putative nova grains in the literature and from this study are compared to the bulk composition of pure CO and ONe nova ejecta. The model predictions are shown as mixing lines between pure nova ejecta and the solar material. The isotope data are plotted with 1σ uncertainties. For each nova model, the number in parenthesis refers to the percentage of H-rich material accreted onto the WD.

Figure 8 Si- and S-isotopic compositions of putative nova and AB grains are compared to model predictions for the He/C zones of different models (solid and dashed lines) and for the C/Si zone of the 25T model (blue circles) by P15. The dotted dashed lines are mixing lines between the He/C and Si/C zones, and the shaded areas therefore represent isotopic compositions that can be produced by local mixing in the 25T model.

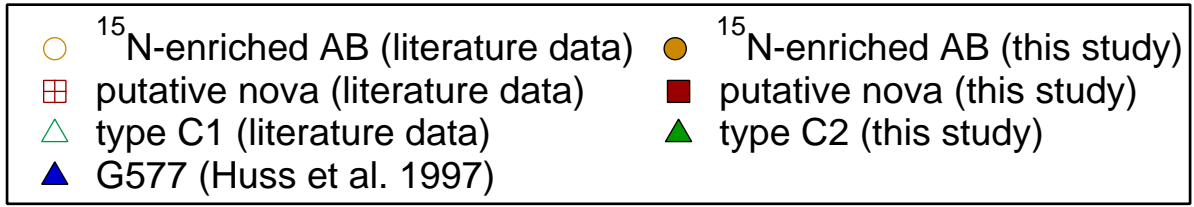
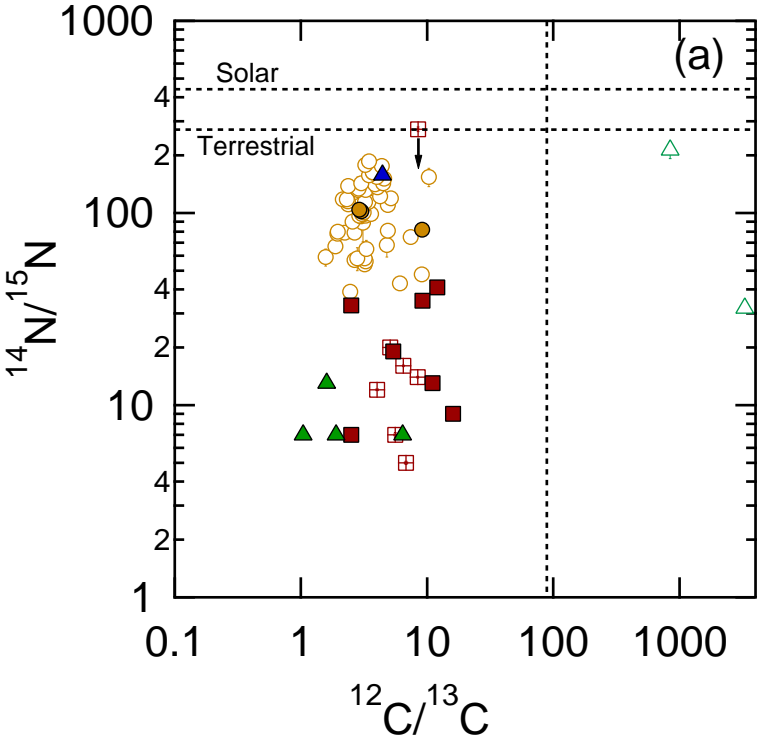
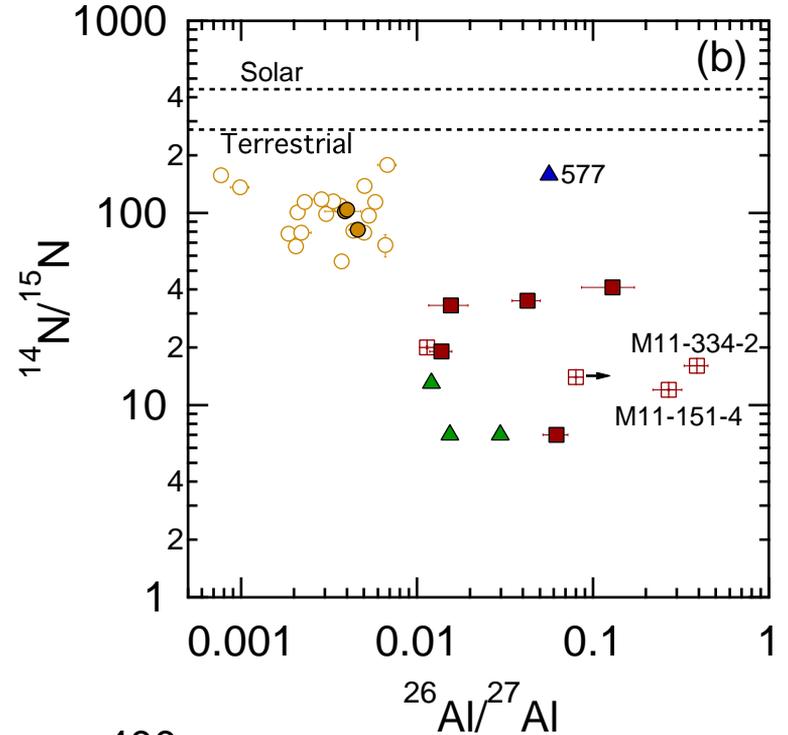
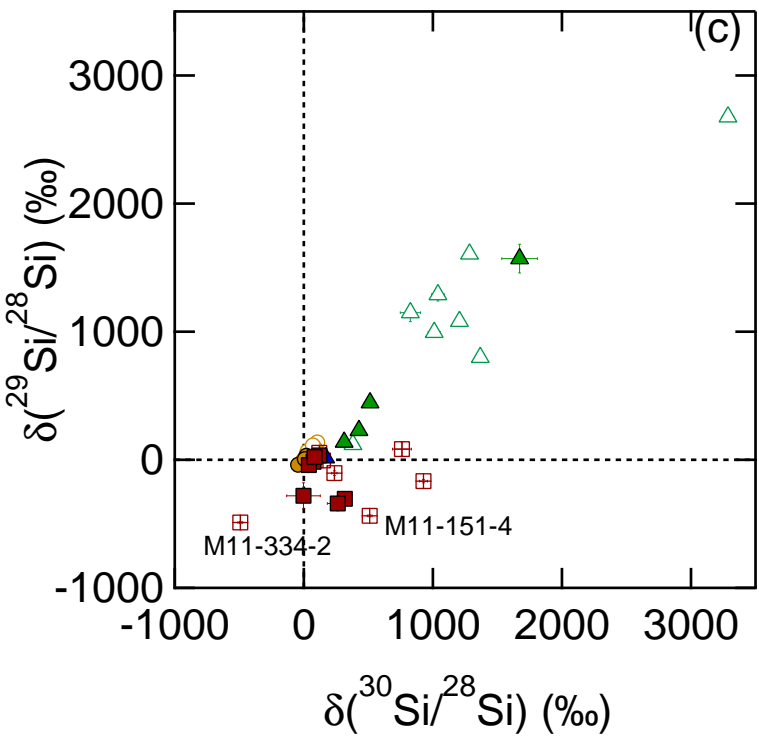
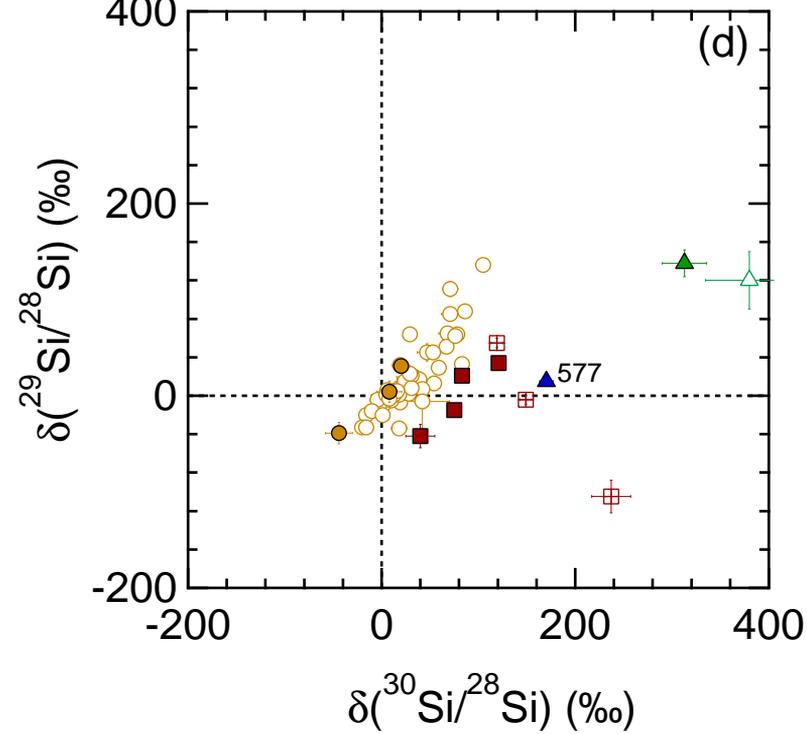

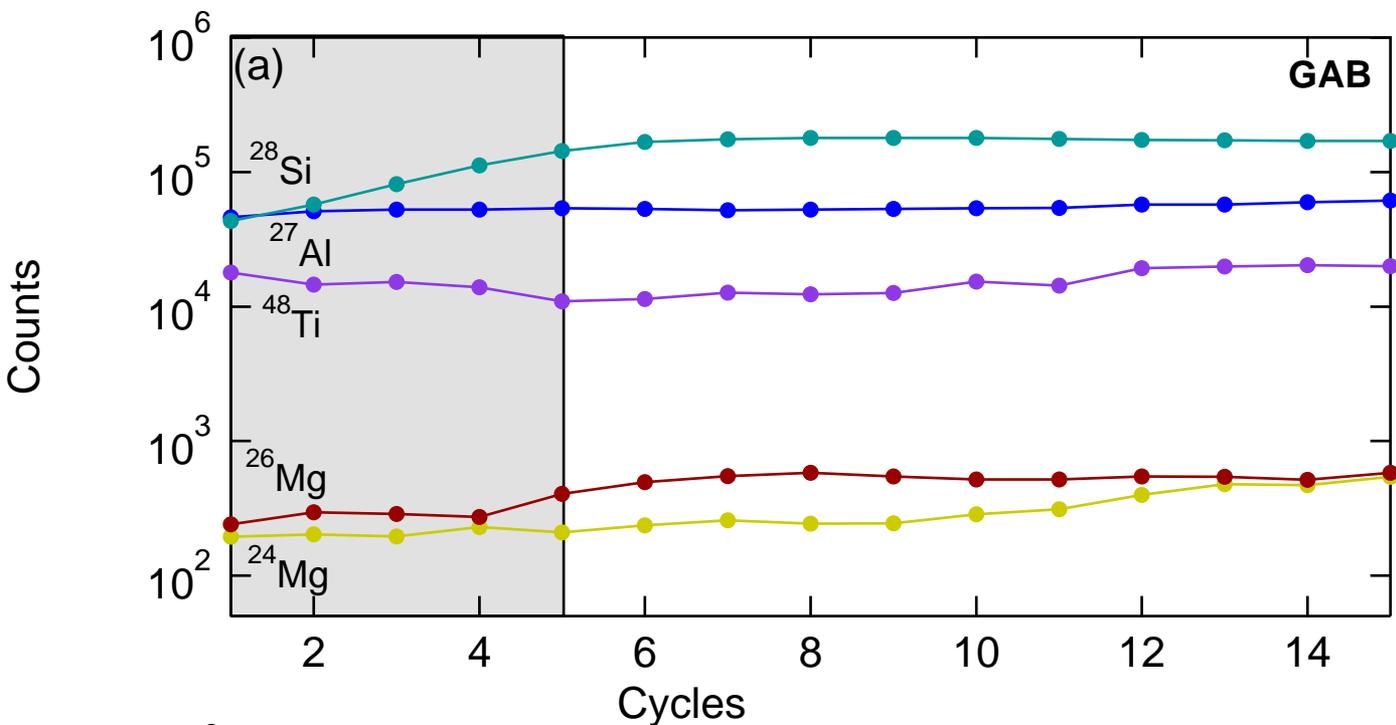
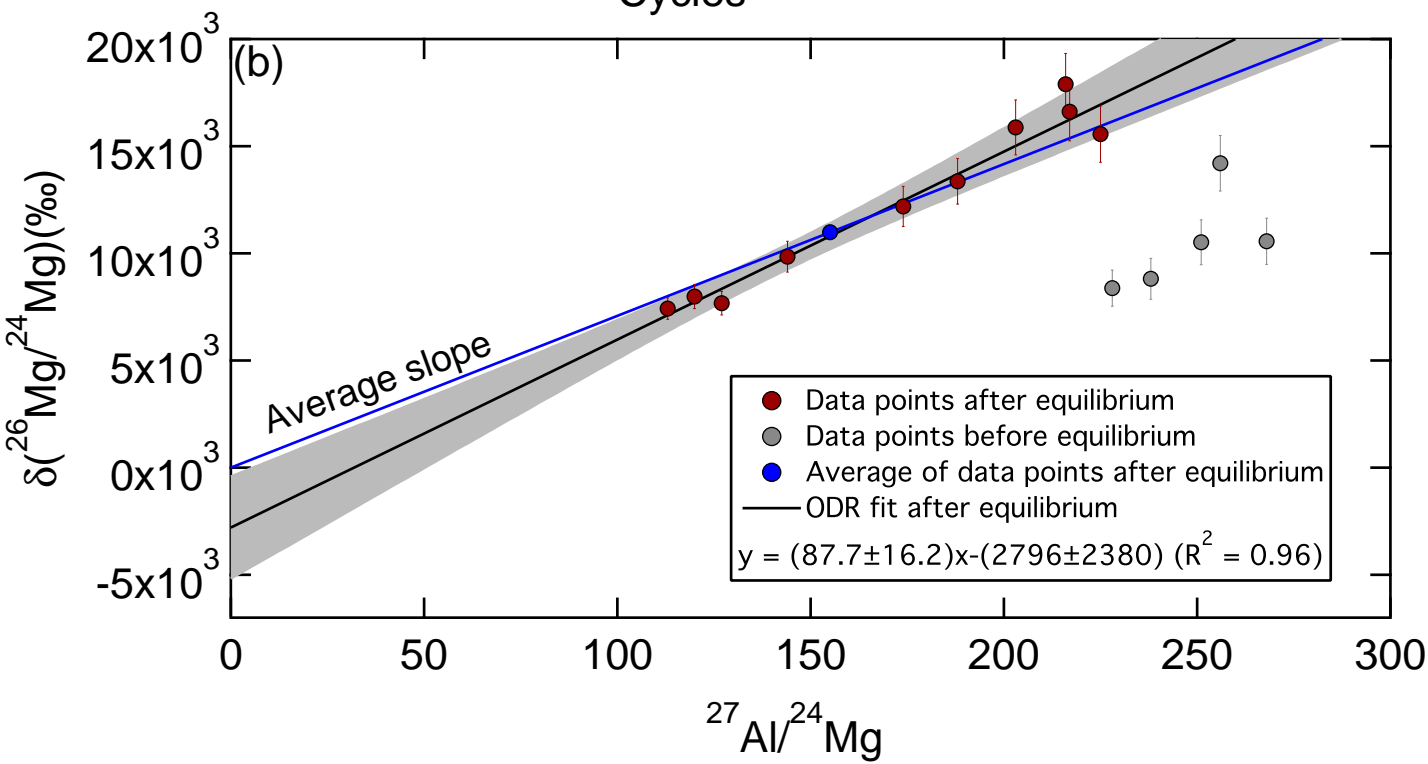

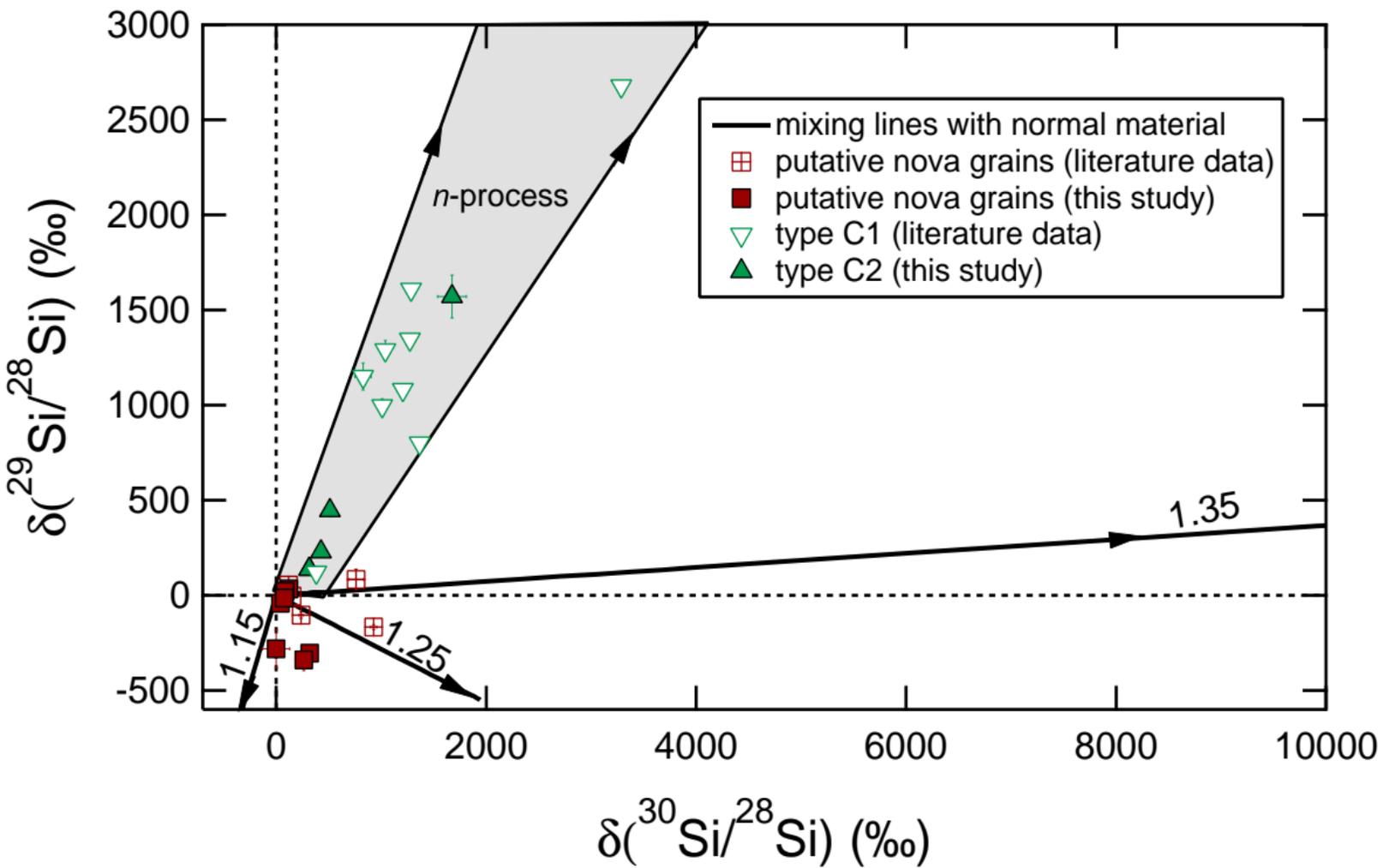

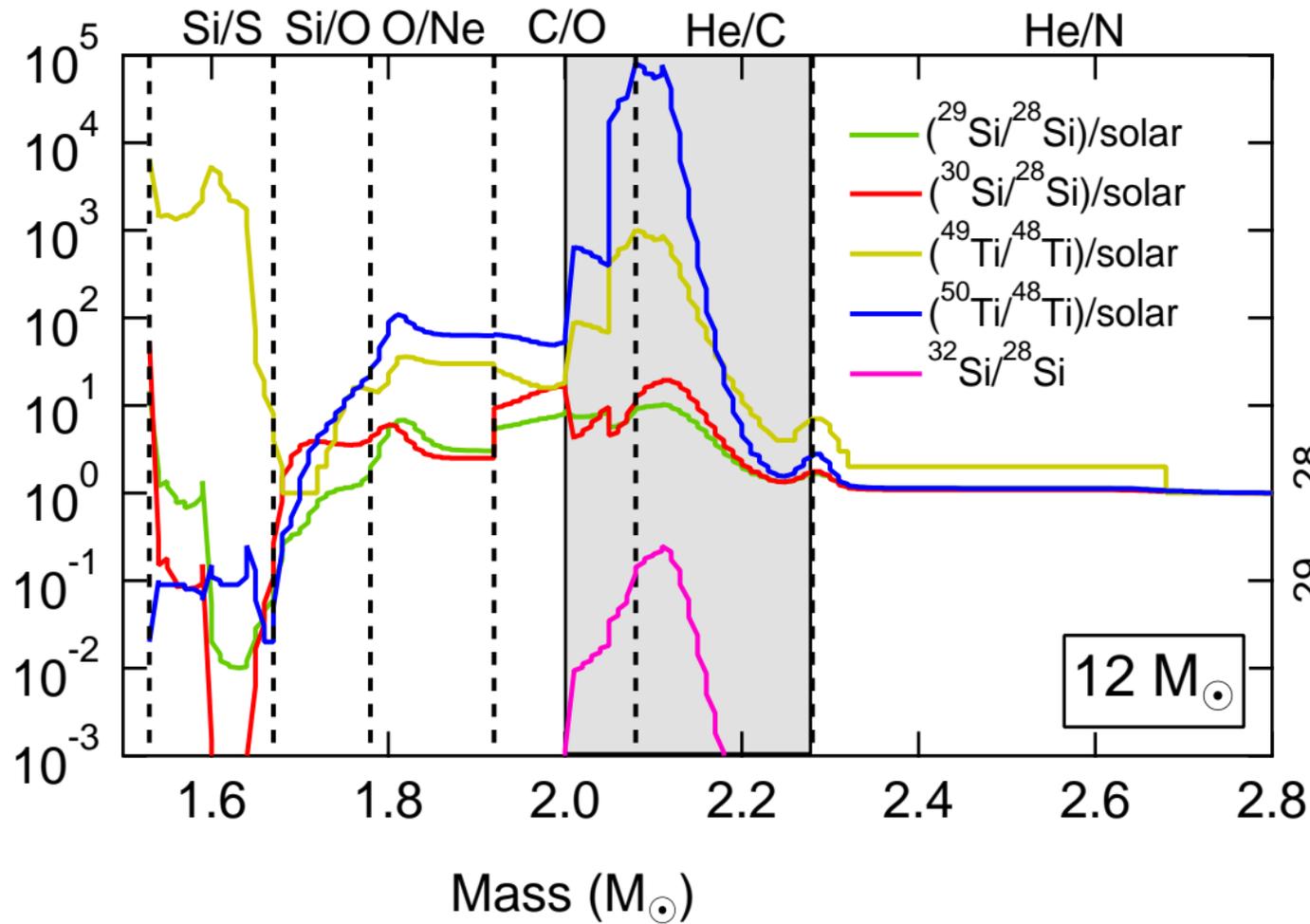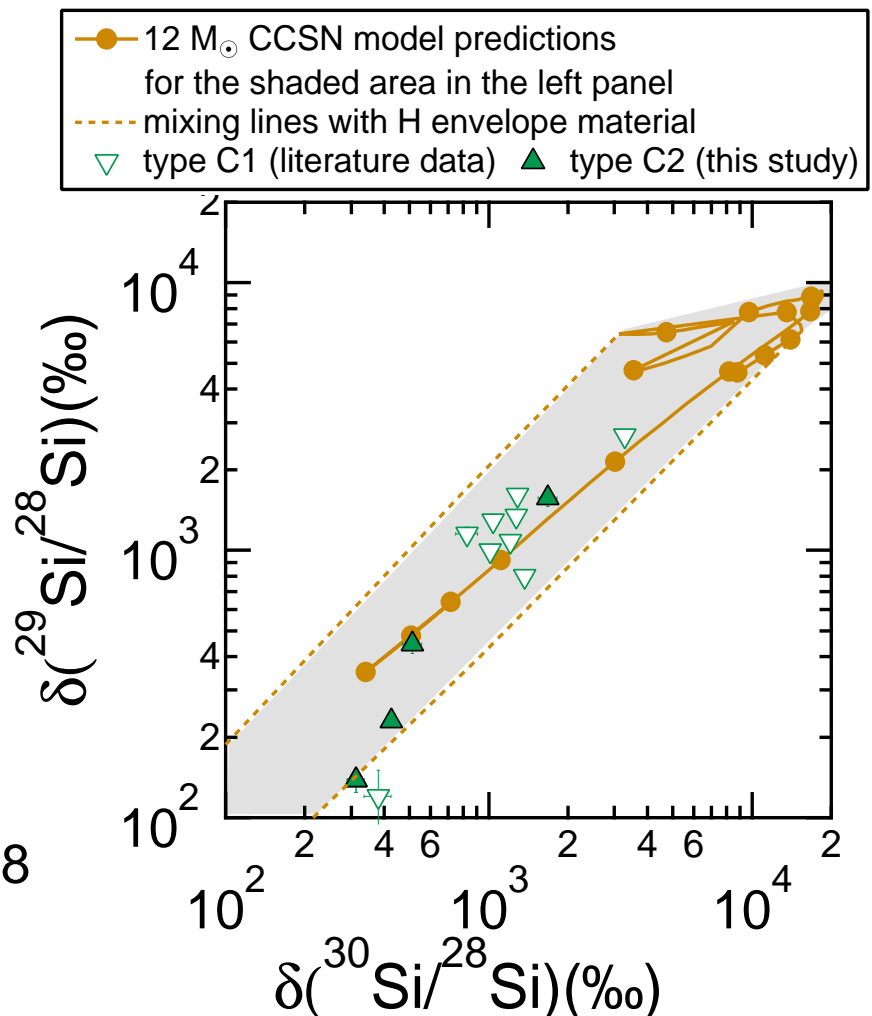

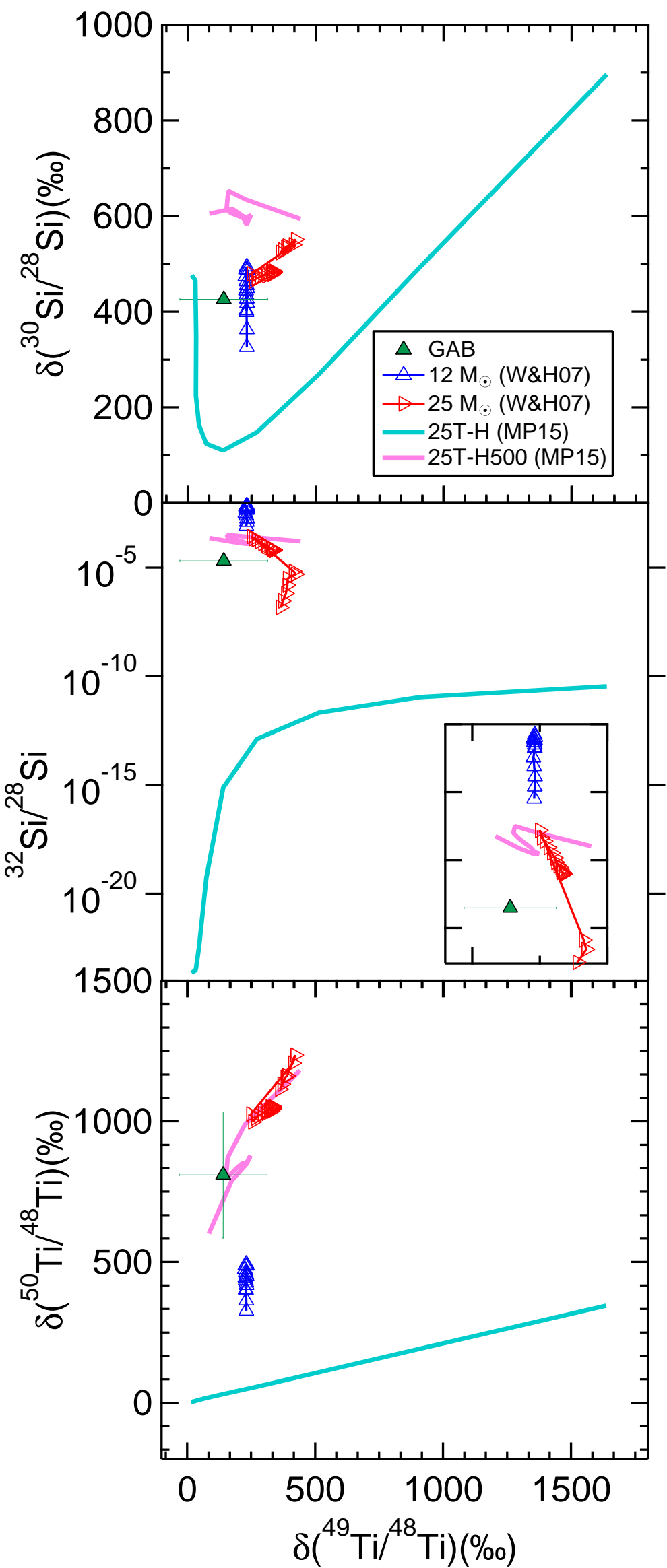

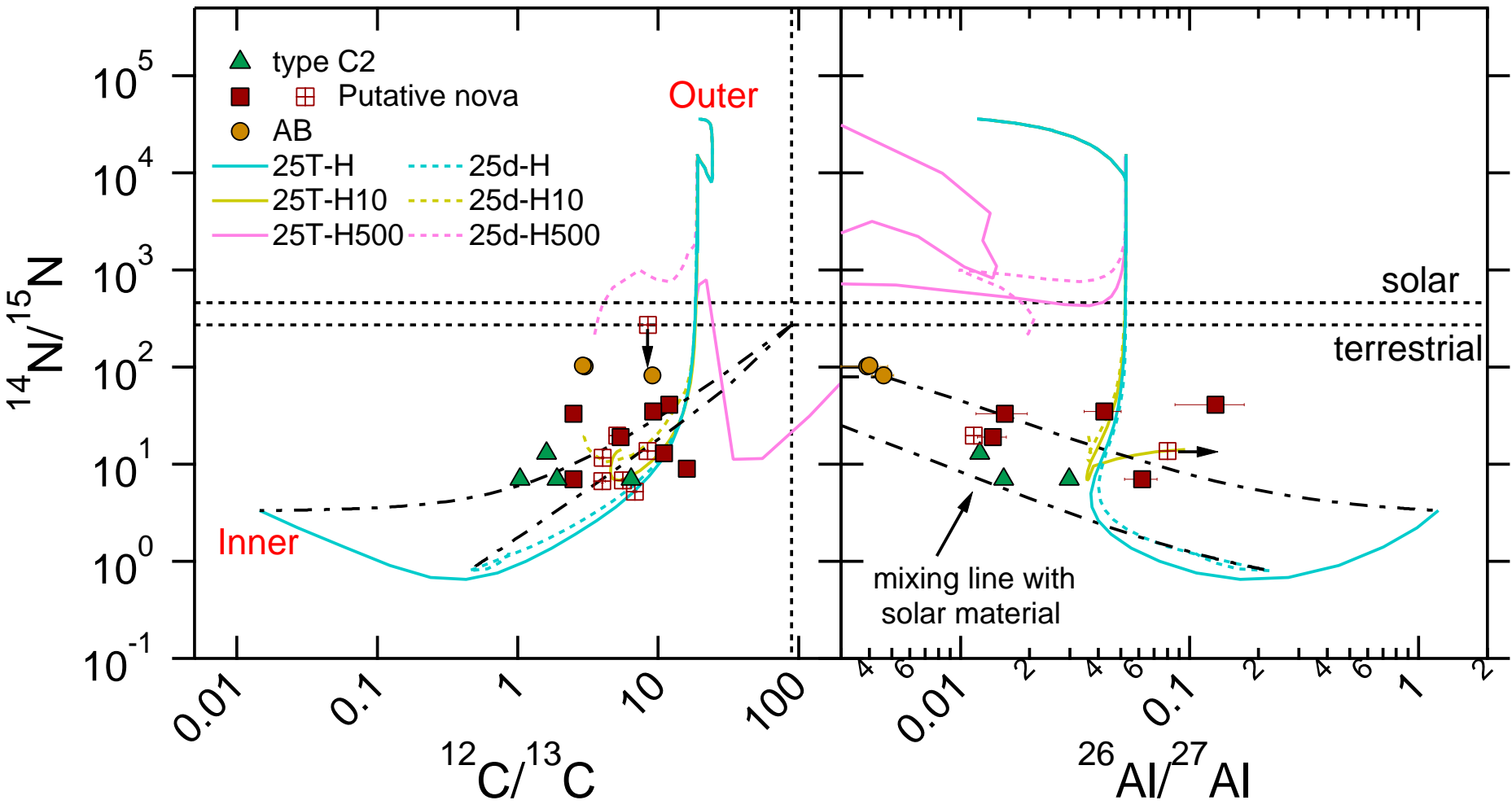

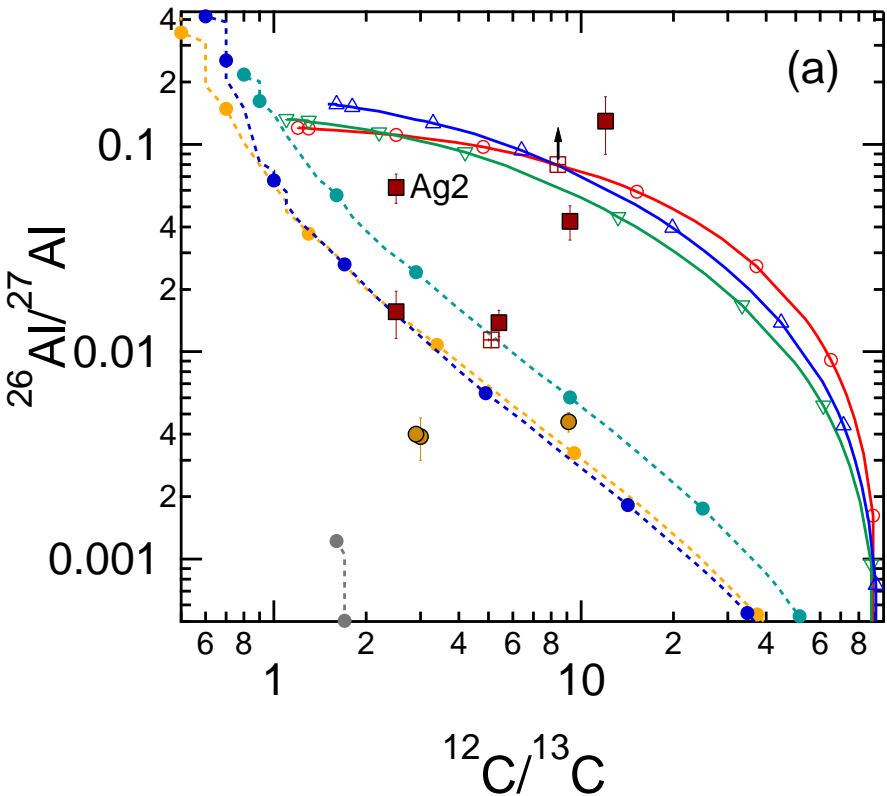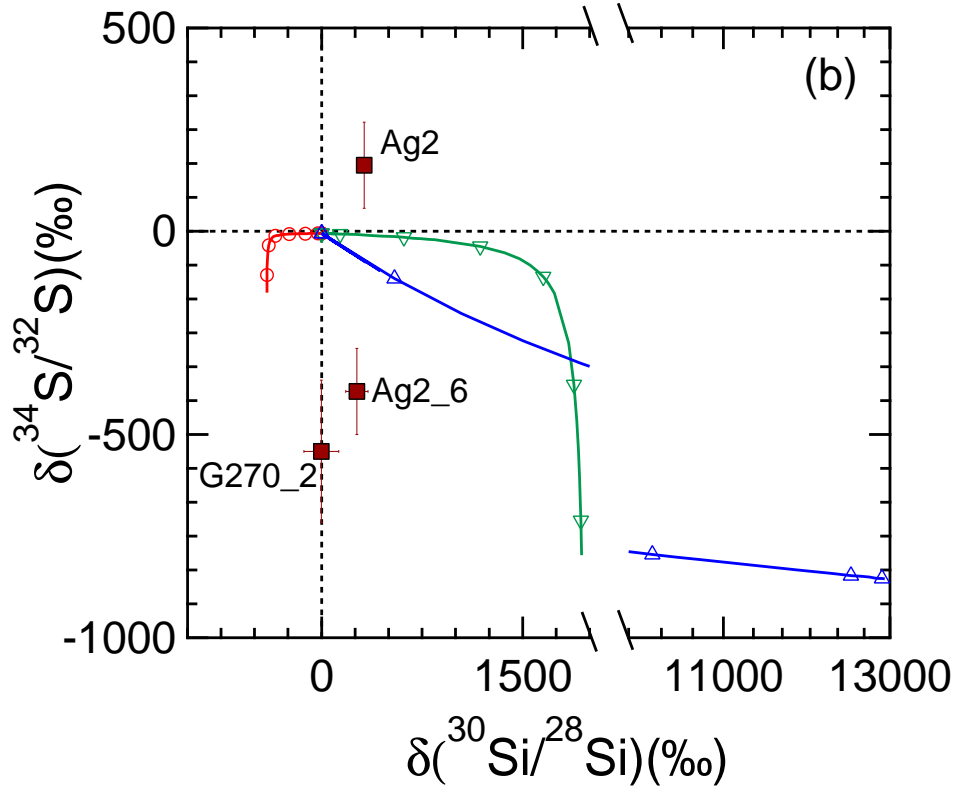

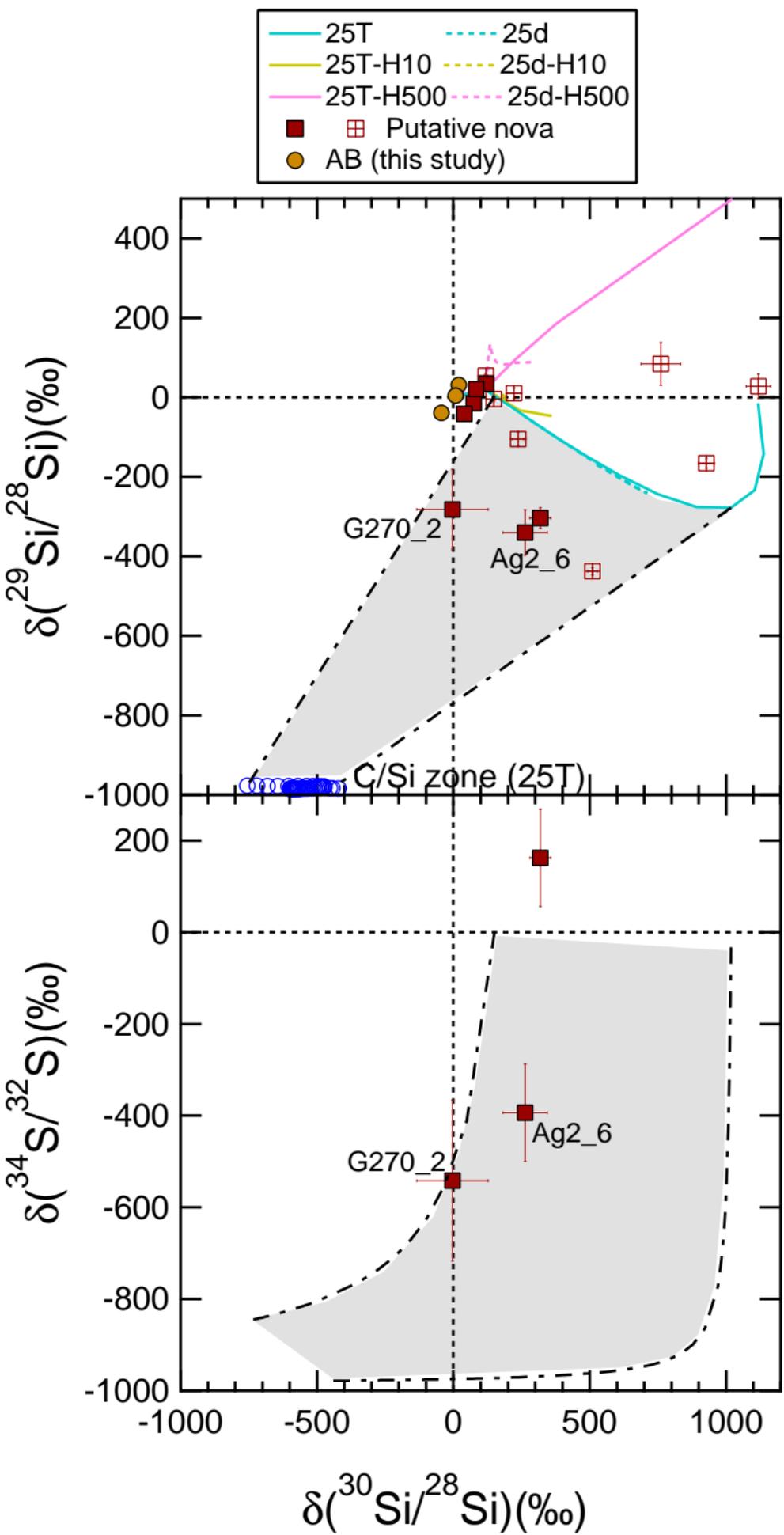

Table 1 Isotope Ratios of $^{13}$C- and $^{15}$N-enriched Presolar SiC Grains*

| Grains | Type | $^{12}$C/$^{13}$C | $^{14}$N/$^{15}$N | δ($^{29}$Si/$^{28}$Si) (‰) | δ($^{30}$Si/$^{28}$Si) (‰) | δ($^{26}$Mg/$^{24}$Mg) (×10$^3$ ‰) | $^{26}$Al/$^{27}$Al[a] (×10$^{-3}$) |
|---|---|---|---|---|---|---|---|
| G278 | C2 | 1.9±0.03 | 7±0.2 | 1570±112 | 1673±138 | | |
| G1342 | C2 | 6.4±0.08 | 7±0.14 | 445±34 | 513±43 | 76±19 | 15.4±0.5 |
| GAB | C2 | 1.6±0.02 | 13±0.2 | 230±6 | 426±7 | 16±0.9 | 12.1±0.8 |
| G240-1[b] | C2 | 1.0±0.01 | 7±0.1 | 138±14 | 313±23 | 960±280 | 29.8±1.1 |
| G270_2 | N | 11±0.3 | 13±0.3 | −282±101 | −3±131 | | |
| Ag2 | N | 2.5±0.1 | 7±0.1 | −304±26 | 319±38 | 315±37 | 62.0±10.0 |
| Ag2_6 | N | 16±0.4 | 9±0.1 | −340±57 | 263±82 | | |
| G283 | N | 12±0.1 | 41±0.5 | −15±3 | 75±4 | 1913±454 | 129.5±43.2 |
| G1614 | N | 9.2±0.07 | 35±0.7 | 34±5 | 121±6 | 184±24 | 42.5±7.8 |
| G1697 | N | 2.5±0.01 | 33±0.8 | −42±12 | 40±15 | 41±8 | 15.6±3.9 |
| G1748 | N | 5.4±0.02 | 19±0.2 | 21±4 | 83±5 | 150±16 | 24.0±4.6 |
| G1485 | AB | 3.0±0.01 | 102±3 | −39±11 | −44±14 | 22±4 | 3.9±0.9 |
| G1516 | AB | 2.9±0.01 | 104±3 | 31±7 | 20±9 | 1±0.05 | 4.0±0.2 |
| G1571 | AB | 9.1±0.04 | 82±3 | 4±11 | 8±13 | 10±0.35 | 4.6±0.5 |

| Grains | δ($^{33}$S/$^{32}$S) (‰) | δ($^{34}$S/$^{32}$S) (‰) | δ($^{44}$Ca/$^{40}$Ca) (‰) | δ($^{47}$Ti/$^{48}$Ti) (‰) | δ($^{49}$Ti/$^{48}$Ti) (‰) | δ($^{50}$Ti/$^{48}$Ti) (‰) | $^{32}$Si/$^{28}$Si[c] |
|---|---|---|---|---|---|---|---|
| GAB | −82±279 | −6±122 | 50±126 | −127±130 | 141±171 | 809±224 | 1.98×10$^{-5}$ |
| G270_2 | −615±385 | −542±175 | | | | | 1.04×10$^{-3}$ |
| Ag2 | −92±222 | 162±106 | | | | | |
| Ag2_6 | 48±334 | −394±106 | | | | | 1.38×10$^{-4}$ |
| G283 | | | −83±61 | 152±76 | −22±70 | 130±77 | |
| G1614 | | | 17±41 | −24±44 | 59±22 | −58±32 | |
| G1697 | | | 221±113 | −29±60 | 35±56 | 33±59 | |
| G1748 | | | 116±94 | −46±44 | 65±25 | 5±36 | |
| G1516 | | | 221±113 | −29±60 | 35±56 | 33±59 | |
| G1571 | | | 116±94 | −46±44 | 64±25 | 5±36 | |

Note: * All isotope data are reported with 1σ uncertainties;

[a] The $^{26}$Al/$^{27}$Al ratio is calculated based on the equation (2) given by Nittler et al. (1997): $^{26}$Al/$^{27}$Al = [($^{26}$Mg/$^{24}$Mg)$_{grain}$−($^{26}$Mg/$^{24}$Mg)$_{std}$]/[($^{27}$Al$^+$/$^{24}$Mg$^+$)×Λ], where Λ is the ratio of Al to Mg sensitivity factors. Because of the probability of Al contamination, the values should be considered as lower limits (see texts in Section 3.2);

[b] Grain G240-1 was previously reported by Nittler et al. (2006);

[c] The $^{32}$Si/$^{28}$Si ratio is calculated from the equation given by Pignatari et al. (2013a): $^{32}$Si/$^{28}$Si = −0.001×$^{32}$S/$^{28}$Si×δS, where δS is the average of δ($^{33}$S/$^{32}$S) and δ($^{34}$S/$^{32}$S). The sensitivity factor ratio of S to Si is three for NanoSIMS measurements (Hoppe et al. 2012). Because S isotopes were simultaneously measured with C and N isotopes in grain G270_2, Ag2, and Ag2_6, the sensitivity factor ratio of Si to C$_2$ is needed to calculate $^{32}$Si/$^{28}$Si (Λ(Si/C$_2$) ~unity according to the ratio determined in both synthetic and presolar SiC).